\documentclass[review]{elsarticle}

\usepackage{url,graphicx,epsfig,graphics,subfigure,psfrag,amsmath,amssymb,epstopdf,enumerate,lineno,hyperref}
\modulolinenumbers[5]

\usepackage{diagbox}
\usepackage{xcolor}
\usepackage[normalem]{ulem}
\newcommand{\stkout}[1]{\ifmmode\text{\sout{\ensuremath{#1}}}\else\sout{#1}\fi}

\journal{Physica A-Statistical Mechanics and Its Applications}

\begin{document}

\begin{frontmatter}

\title{Study of phase transition of Potts model with Domain Adversarial Neural Network}
%
\author[mymainaddress]{Xiangna Chen}

\author[mymainaddress,secondaryaddress]{Feiyi Liu\corref{mycorrespondingauthor}}
\cortext[mycorrespondingauthor]{Corresponding author}
\ead{fyliu@mails.ccnu.edu.cn}

\author[mymainaddress]{Shiyang Chen}
\author[Fourthaddress]{Jianmin Shen}
\author[mymainaddress]{Weibing Deng}\ead{wdeng@mail.ccnu.edu.cn}
\author[secondaryaddress]{G\'abor Papp}
\author[mymainaddress]{Wei Li}
\author[mymainaddress]{Chunbin Yang}

\address[mymainaddress]{Key Laboratory of Quark and Lepton Physics (MOE) and Institute of Particle Physics, Central China Normal University, Wuhan 430079, China}

\address[secondaryaddress]{Institute for Physics, E{\"o}tv{\"o}s Lor\'and University\\1/A P\'azm\'any P. S\'et\'any, H-1117, Budapest, Hungary}


\address[Fourthaddress]{School of Engineering and Technology, Baoshan University,  Baoshan 678000, China}

\begin{abstract}

A transfer learning method, Domain Adversarial Neural Network (DANN), is introduced to study the phase transition of two-dimensional $q$-state Potts model.
With the DANN, we only need to choose a few labeled configurations automatically as input data, then the critical points can be obtained after training the algorithm.
By an additional iterative process, the critical points can be captured to comparable accuracy to Monte Carlo simulations as we demonstrate it for $q = 3,4, 5, 7$ and $10$.
The type of phase transition (first or second-order) is also determined at the same time. Meanwhile, for the second-order phase transition at $q=3$, we can calculate the critical exponent $\nu$ by data collapse. Furthermore, compared to the traditional supervised learning, we found the DANN 
to be more accurate with lower cost.

\end{abstract}


\end{frontmatter}


\section{Introduction}
\label{intro}

As an important theme of artificial intelligence, machine learning (ML) has been successfully applied in various fields of science and technology \cite{jordan2015machine}, such as speech recognition \cite{hinton2012deep},  image classification  \cite{krizhevsky2012imagenet,obuchi2016boltzmann}, vehicle autopilot \cite{he2020machine,stilgoe2018machine}, protein folding \cite{noe2020machine,xu2019distance}, fusion reactor \cite{morgan2022machine}, etc.
As its competency to capture features and classifications, ML has also been widely employed to data processing in statistical physics \cite{engel2001statistical,mehta2014exact,mehta2019high,carleo2019machine,carrasquilla2020machine}, especially as a tool to study the phase transition in many-body systems \cite{carleo2017solving}, the topological phases of matter \cite{deng2017machine,carrasquilla2017neural} and so on.

In recent years, the commonly used learning methods to study phase transitions are either supervised learning \cite{carrasquilla2017machine,van2017learning,
zhang2019machine,tanaka2017detection,
tomita2020machine,li2018applications,
yau2022generalizability,
tan2021universal,tan2020comprehensive} or unsupervised learning \cite{wang2016discovering,shen2021supervised,
wetzel2017unsupervised,hu2017discovering,
wang2017machine,wang2021unsupervised,
giataganas2022neural,zhang2022machine}.
While supervised learning requires labeled data as the input and is mainly used to identify or classify the phases of matter,
unsupervised learning does not need any labelling
of the order parameters.  
The later is more adequate for clustering, and dimensionality reduction 
using e.g. principal component analysis (PCA)  \cite{wang2016discovering,wetzel2017unsupervised,abdi2010principal},  t-distributed stochastic neighbor embedding (T-SNE)  \cite{ch2018unsupervised,van2014accelerating,wattenberg2016use} or nonlinear autoencoder (AE)  \cite{wetzel2017unsupervised,ch2018unsupervised}.   
However, both the supervised learning and unsupervised learning have good performances in multiple types of models \cite{carrasquilla2017machine,van2017learning,
shen2021supervised,wattenberg2016use}.
 
Although the above methods possess their own advantages, each has its limitations, which include, for example, the time-consuming nature of supervised learning, for the fact that labeled data is not readily available; the incompleteness of unsupervised learning (the dynamic features of the data cannot be fully identified, or more data samples are required), these factors are undoubtedly very important. In view of this, transfer learning (TL) \cite{pan2009survey,weiss2016survey}, a (semi-supervised) method (part of the data is labeled) mixing both labeled and unlabeled data, has been proposed to study the phase transition \cite{ch2017machine,malo2019applications}. Inspired by the idea of translating unlabeled data in target domain into labeled data in source domain, TL can not only obtain the critical exponent of the phase transition model through data collapse as supervised learning does, but also extract the feature representation from the original data as unsupervised learning does. 

In this paper, a specific technique of TL, the Domain Adversarial Neural Network (DANN) \cite{ajakan2014domain,ganin2016domain}, is introduced to investigate phase transition. DANN is a network approach based on the domain adaptation method (DA), the idea of which is to ensure that the source domain (labeled data) and the target domain (usually no labels or only very few labels) share the same features, and improve the performance of the model in the target domain by training \cite{ben2006analysis,farahani2021brief}.
More specifically, this property allows us to predict the critical point when it is unknown.
DANN has been successfully applied to detect phase transition and predict critical point of Bose-Hubbard model and the Su-Schrieffer-Heeger model with disorder, and also analyze the many-body localization problem  \cite{huembeli2018identifying}. Through the study of non-equilibrium and equilibrium phase transition models, such as direct percolation (DP) and two-dimensional site percolation \cite{shen2021transfer}, DANN is also shown to outperform the traditional supervised learning as respects to the efficiency and accuracy.

To demonstrate the robustness of the DANN and its suitability for handling more complex phase transition models, here we employed the DANN to investigate the phase transition of the two-dimensional $q$-state Potts models on the square lattice with nearest-neighbor interactions. 
 One of the important characteristics of the Potts model is that, the first or second order phase transition depends on the number of states $q$. 
Therefore,  we would test the power of DANN on predicting the first or second order phase transition.
Our results suggest that  DANN 
is applicable for  classifying phase transitions and predicting the critical points of the two-dimensional $q$-state Potts model. 

The rest of this paper is organized as follows. In Section \ref{model}, we introduced the $q$-state Potts  model and the Monte Carlo Glauber algorithm for Potts model. The DANN method, the data sets of the model and the optimal source domain of DANN are presented in Section \ref{method}. In Section \ref{sec:Results}, we provide the various results and analysis of the DANN. Finally, the last Section is devoted to the conclusion.

\section{Model}
\label{model}
\subsection{ The q-state Potts model}
\label{potts}

The $q$-state Potts model is a classical model of spin systems which describes the classical spin interactions on the lattice \cite{potts1952some,wu1982potts}. 
The Hamiltonian of the $q$-state Potts model is as follows:
\begin{equation}
\mathcal{H}=-J \sum _{\langle i,j \rangle}  \delta_{\sigma_i,\sigma_j}, \quad \sigma_i,\sigma_j\in \{0, 1,..., q-1\},
\end{equation}
where $\sigma_i$ is the spin  value at site $i$. $\langle i,j \rangle$ represents the sum of nearest neighbor pairs over all lattice sites. $\delta$ is the Kronecker delta. $J$ is defined as the interaction constant. At the thermodynamic limit ($L \rightarrow\infty $), the critical temperature $T_c$ satisfies   \cite{potts1952some,den1979relation}:
\begin{equation}
T_{c}=T _{c,q}=\frac{1}{\ln(1+\sqrt{q})}{J}/{k_B},
\label{eq:Tc-theor}
\end{equation}
where $k_B$ is the Boltzmann constant.

In two-dimensional system, the $q$-state Potts model can be ferromagnetic or non-ferromagnetic, and its phase transition properties are closely related to the spatial dimension\;\cite{baxter2016exactly}.
The situation of $J > 0$ corresponds to ferromagnetic Potts model, which means the spins tend to be aligned  in this nearest-neighbor interaction; $J < 0$  corresponds to anti-ferromagnetic; and the spins are non-interacting if $J = 0$. Here we use the ordinary ferromagnetic Potts model with $q$-valued spins considering only nearest-neighbor interactions ($J>0$) and the sum is running on a square lattice $L \times L$ with periodic boundary conditions. To study the phase transition behavior, we set the constant $J/k_{B}=1$. It should be also emphasized that the 2D (ferromagnetic) Potts model exhibits a second-order phase transition for $q \leq 4$ and a first-order phase transition for $q >4$, corresponding to continuous and discontinuous transitions at the critical temperature $T_c$ \cite{wu1982potts,baxter1973potts}, respectively. These are what we intend to distinguish solely through the power of DANN.

\subsection{Monte Carlo  Glauber Algorithm for Potts Model}\label{Glauber}

For the two-dimensional Potts model with periodic boundary
conditions, the spin value $\sigma_i$ of a site $i$ being occupied in the next Monte-Carlo step is controlled by the flip probability $p$, so we describe the spin-hopping (spin-flipping) mechanism as follows,
\begin{equation}
  \sigma_i=\left\{
\begin{array}{lll}
\sigma_i^{in}     &      & \text{if} \quad r>p, \quad \text{flip is not accepted}, \\
\sigma_i^{out}     &      & \text{if}  \quad r\leq p, \quad \text{flip is accepted}, 
\end{array} \right.
\end{equation}
where $r$ is a random number in $(0, 1)$.
$\sigma_i^{in}$ and $\sigma_i^{out}$ are the spin values of site $i$ before and after flipping, and $\sigma_i^{out}$ is chosen randomly from the remaining states.

The flip probability $p$ is related to the spin values of the neighbor sites.
The algorithm performs configuration evolution  according to the dynamic rules \cite{henkel2008non,glauber1963time,mariz1990comparative}:

\begin{equation}
p=\frac{1}{1+exp(\Delta \epsilon *\beta)}, \quad  \beta=\frac{1}{ T},
\end{equation}
where the  energy $ \epsilon$ of site $i$ is:
\begin{equation}
 \epsilon =  -(\delta_{\sigma_i,\sigma_{j-1}}+\delta_{\sigma_i,\sigma_{j+1}} +\delta_{\sigma_{i-1},\sigma_j}+\delta_{\sigma_{i+1},\sigma_j}).
 \label{eq:energy}
\end{equation}
Here $i \pm 1,j \pm 1$ indicate the horizontal and vertical nearest neighbour sites (two adjacent sites) , and
 $\Delta \epsilon =\epsilon^{out}-\epsilon^{in}$ is the energy difference  before and after the flip of the site $i$.
The flip probability $p$, has a temperature dependence, allowing to obtain different phases, when the equilibrium is reached.
   
\begin{figure}[htbp]
\centering
\subfigure[$q=3,T=0.05<T_c$]{
\label{Fig.configq3.1}
\includegraphics[width=0.3\textwidth]{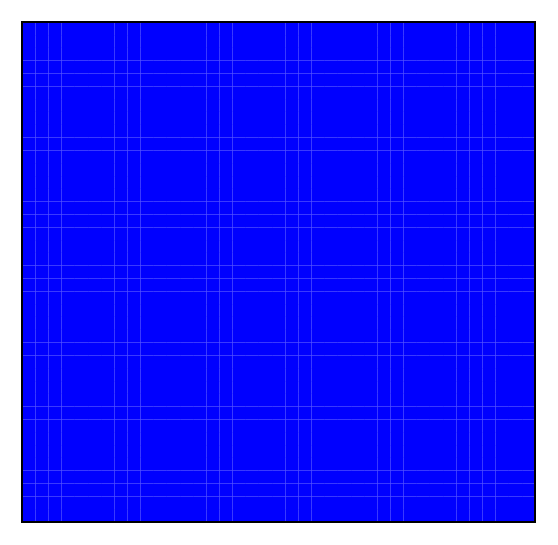}}
\subfigure[$q=3,T=0.9949 = T_c$]{
\label{Fig.configq3.2}
\includegraphics[width=0.3\textwidth]{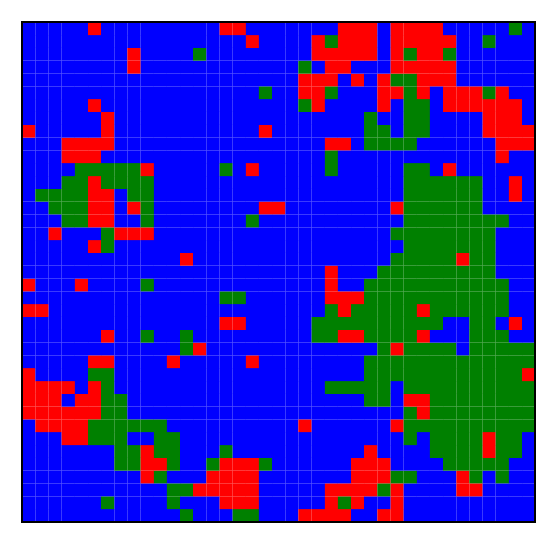}}
\subfigure[$q=3,T=4.0>T_c$]{
\label{Fig.configq3.3}
\includegraphics[width=0.3\textwidth]{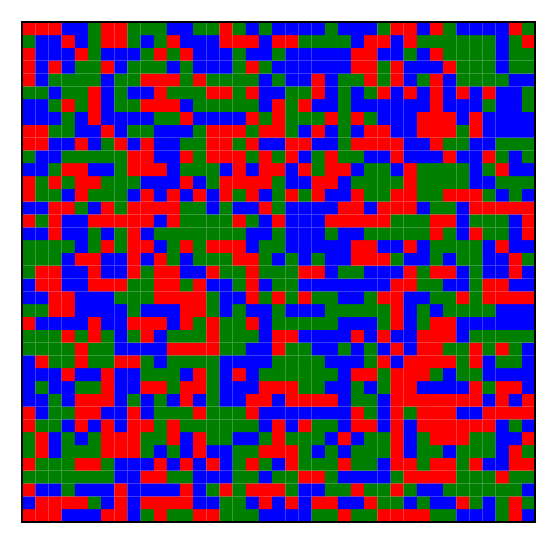}}\\

\caption{The two-dimensional Potts model generated by Monte Carlo simulation with $q = 3$ at $L = 40$, after  $100000$ time steps. (a) $T=0.05 < T_c$; (b) $T =T_c=0.9949$; 
(c) $T=4.0 > T_c$. Note the flip of the central blue cluster to 
other colors. }
\label{fig:config}
\end{figure}

\begin{figure}[htbp]
\centering
\subfigure[$q=10,T=0.05<T_c$]{
\label{Fig.configq3.1}
\includegraphics[width=0.3\textwidth]{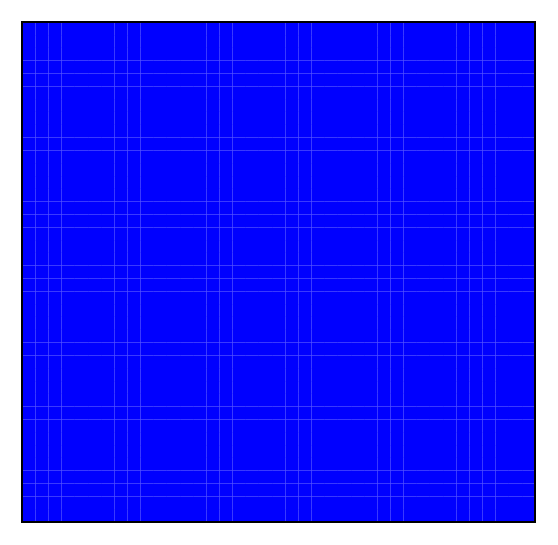}}
\subfigure[$q=10,T=0.7012=T_c$]{
\label{Fig.configq10.2}
\includegraphics[width=0.3\textwidth]{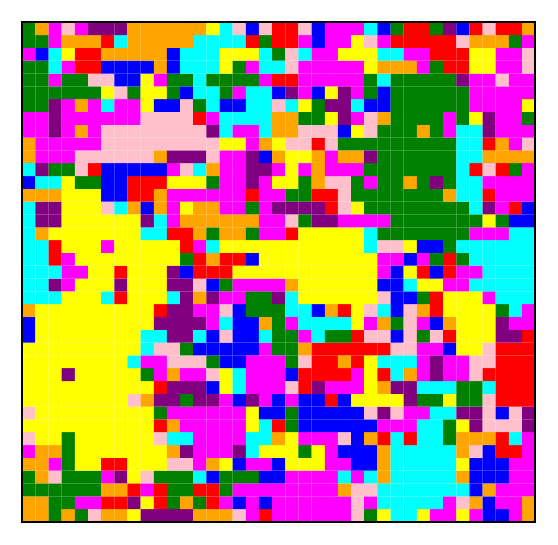}}
\subfigure[$q=10,T=4.0>T_c$]{
\label{Fig.configq10.3}
\includegraphics[width=0.3\textwidth]{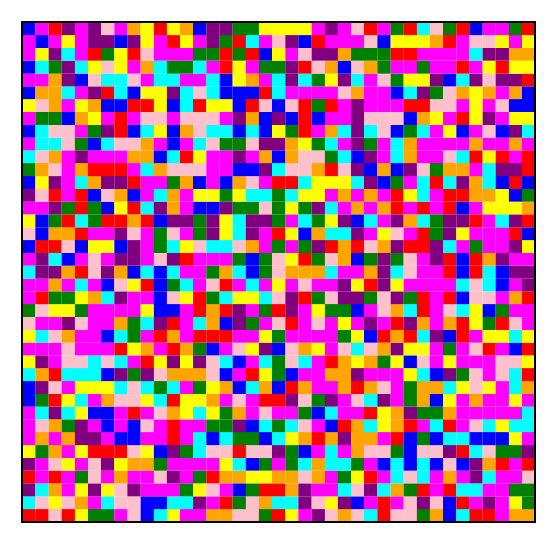}}

\caption{The  spin configurations (snapshots) of Potts model with $ q=10$. }
\label{fig:config_q510}
\end{figure}

Fig.~\ref{fig:config} is an instantaneous snapshot of the $3$-state 
Potts model configuration after $t=100000$
Monte Carlo time steps 
at lattice size $L=40$,
for different temperatures across the critical one, indicating the change of phases in a fully occupied lattice.
For $T=0.05<T_c$, the configuration of $q=3$ is an ordered state after reaching equilibrium.
For $T=T_c=0.9949$, one integer value (one type of color dominates, such as blue) occupies most of the lattice
with small clusters of the other spin directions.
If $T$ is large enough ($T=4.0>T_c$),  countless small clusters will occupy the system, which means the system becomes disordered. 
This feature remains at higher $q$ values, e.g. the $q=10$ Potts model 
 also shows a
characteristic of disorder at high temperature and order at low temperature (see Fig.~\ref{fig:config_q510}).

As input of DANN, these configurations are stored for  
$125$ temperature values for $T$ in $[0.05, 4]$ (a large range on  
both sides of the critical temperature $T_c$) for each $q$ with  
the given lattice size $L$, where $q = 3,4,5,7$ and $10$. To assess the effects arising from the finite size, we choose $L = 20, 30, 40, 50$ and $60$. The total simulation time is set to $t = 100000$ Monte-Carlo steps to ensure that the system has reached equilibrium.
Usually, to maintain sample independence, only one sample at a time is retained when the equilibrium state is reached, but it 
is inefficiency and time consuming.
Therefore, we take another methods instead,
that is, after reaching the equilibrium for the given temperature $T$, to ensure the independence between configurations (samples), a configuration is saved every $200$ Monte Carlo time steps until the total number of samples $n=1000$.
In the paper, we use these configurations to perform the analysis of the phase structure by DANN.

\section{Method}
\label{method}
\subsection{The Domain Adversarial Neural Network (DANN) method}\label{dann}
By feeding the configurations to DANN as input data, the training process of network follows. 
For transfer learning, these input configurations need to be partly labeled, we will introduce the labeling rules in detail in Sec. \ref{dataset}. One item we would like to point out is that, for each $q$ state of the data we train the network separately, although the architecture of the network is the same.

\begin{figure}[htbp]
\centering
\includegraphics[width=1\textwidth]{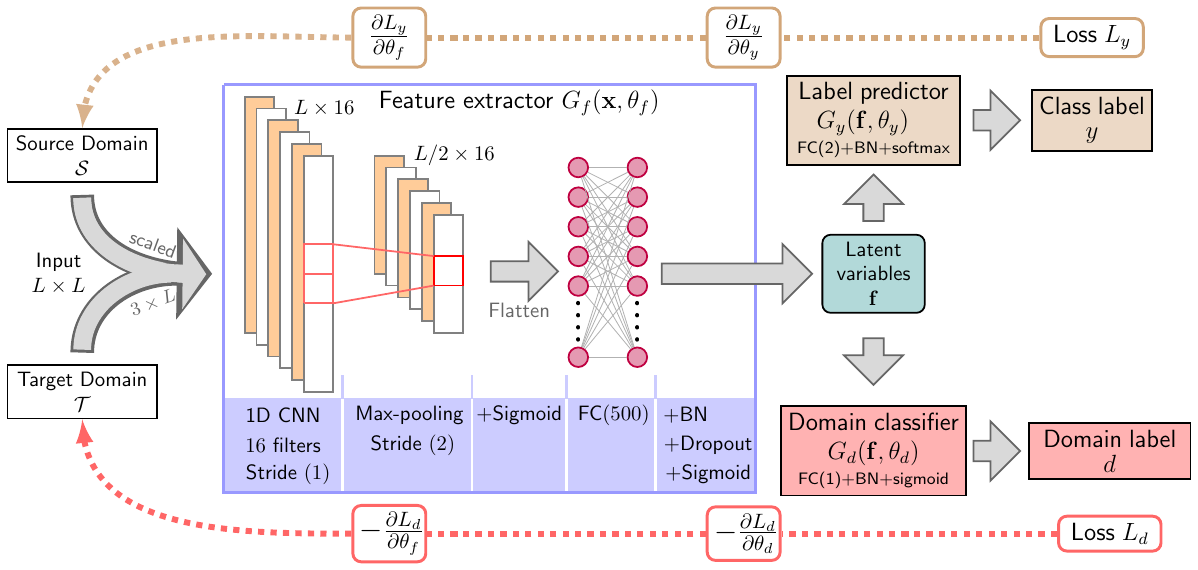}
\caption{Schematic representation of the entire operating framework of DANN. The source domain $\mathcal{S}$ refers to labeled data,  the target domain $\mathcal{T}$ refers to unlabeled data.
DANN mainly consists of three parts: feature extractor, label predictor and  domain classifier.
The feature extractor is used to generate feature vector (Latent variables), 
which is  fed to label predictor and domain classifier to predict labels and distinguish  whether the data is coming from the source or target domain, respectively.
These three parts can be realized by training three corresponding parameters $\theta _f$,  $\theta_y$, $\theta _d$. $L$ stands for loss.}
\label{fig:DANNtk}
\end{figure}

The process of our DANN algorithm is shown in Fig.~\ref{fig:DANNtk}, which mainly consists of three parts: feature extractor, label predictor and domain classifier. The input variable $\mathbf{x}$ is divided into two distribution of groups: 
the source domain $\mathcal{S}=\{(x_s, y_s)\}$ with labeled data $x_s$ and its label $y_s$, and the target domain $\mathcal{T}=\{ x_t \}$ with unlabeled data $x_t$, where $\mathbf{x}=\{x_s\}\cup \{x_t\}$. 
The core idea is using the adversarial domain adaptation to achieve the same distribution of data in the source domain $\mathcal{S}$ and target domain $\mathcal{T}$, and predict the corresponding labels for unlabeled data $x_t$~\cite{ben2006analysis,farahani2021brief}.

In the process of training DANN, the first step is to feed the dataset $\mathcal{S}\cup\mathcal{T}$ to feature extractor $G_f(\mathbf{x},\theta _f)$, for mapping to a high-dimensional feature vector $\mathbf{f}= G_f(\mathbf{x}, \theta _f)$ with parameter $\theta _f$. The feature extractor $G_f(\mathbf{x}, \theta _f)$ has the structure of a convolutional neural network (CNN), connected with a fully connected network (FCN) layer, as shown in the largest blue box in Fig.~\ref{fig:DANNtk}. The input configurations $x\in\{x_s\}\cup\{x_t\}$ as images of size $L\times L$ are scaled by a kernel of size $3\times L$ and convoluted into $16$ filters forming feature maps, which grabs the locations and strength of detected features. Then a max-pooling layer is set to reduce the size of those feature maps to $L/2\times16$. With a flattening process, they become a set of distribution in $[ 0,1]$ through the sigmoid activation function. After that, we put the data into a fully connected layer with $500$ neurons to combine the features with a wider variety of attributes. Through an additional process combined with a batch normalization, a dropout (with rate $0.8$) is applied to avoid overfitting problems together with a hard sigmoid map for faster and more stable results, at last we can get the feature vectors $\mathbf{f}$.

Next, the feature vectors, $\mathbf{f}$ are fed into the label predictor, $G_y (\mathbf{f}, \theta _y)$ and domain classifier, $G_d (\mathbf{f}, \theta _d)$ with parameters $\theta _y$ and $\theta _d$. 
The label predictor $G_y (\mathbf{f}, \theta _y)$ consists of $2$ neurons with a batch normalization and the softmax activation function, and the output is a two-dimensional vector whose two elements denote the probabilities of the configurations belonging to category ``0" (ordered)  and category ``1" (disordered), respectively. The softmax activator is to ensure the sum of elements of the vector is always $1$.
The domain classifier $G_d (\mathbf{f}, \theta _d)$ has a structure of $1$ neuron with a batch normalization and hard sigmoid activation function, and outputs the corresponding labels $d$ by identifying whether the feature vector $\mathbf{f}$ is from the source domain (for labeled data $d=0$) or the target domain (for unlabeled data $d=1$).

To predict the label for $x_t$ in target domain, we need to maximize the accuracy (minimize the loss) of the label predictor $G_d (\mathbf{f}, \theta _d)$ so that data from two domains ($d=0$ or $d=1$) cannot be separated by the domain classifier $G_y (\mathbf{f}, \theta _y)$. 
Since the data in the source domain is labeled and can be used as a benchmark data, the loss $L_{y}$ of the labeled predictor is calculated by the feature representation of the source domain.
The loss function $L_{d}$ of the domain classifier is directly designed through the feature representation of the entire dataset. 
As described in Ref. \cite{huembeli2018identifying,shen2021transfer}, the total loss function  can be expressed as follows,
\begin{equation}
L(\theta _f,\theta _y,\theta _d)=L_y(\theta _f,\theta _y)-L_d(\theta _f,\theta _d).
  \label{loss_Lf}
\end{equation}
The whole training process is to optimize $L$ by transforming each parameter $\theta_f$,  $\theta_y$, $\theta_d$ (finally finding the saddle point $\hat{\theta}_f$, $\hat{\theta}_y$ and $\hat{\theta}_d$):
\begin{align}
\hat{\theta}_f,\hat{\theta}_y =\mathop{\arg\min}\limits_{\theta _f,\theta _y} L(\theta _f,\theta _y, \hat{\theta}_d),\\
\hat{\theta}_d =\mathop{\arg\max}\limits_{\theta _d} L(\hat{\theta}_f,\hat{\theta}_y, \theta_d).	
\label{loss_argmin}
\end{align}
The transformation of $\theta_f$,  $\theta_y$, $\theta_d$ can be implemented by the gradient update, and $\hat{\theta}_f$, $\hat{\theta}_y$ and $\hat{\theta}_d$ are found to be the stationary points. The update rules with learning rate $\mu = 0.0001$ are as follows:
\begin{align}
 \label{loss_f} &\theta_f \quad\leftarrow \quad \theta_f-\mu\left(\frac{\partial L_y}{\partial\theta_f}- \frac{\partial L_d}{\partial\theta_f}\right),\\ 
 \label{loss_y} &\theta_y \quad\leftarrow \quad \theta_y- \mu\left( \frac{\partial L_y}{\partial\theta_y}\right),\\ 
 \label{loss_d} &\theta_d \quad \leftarrow \quad  \theta_d- \mu \left( \frac{\partial L_d}{\partial\theta_d}\right).
 \end{align}
From the opposite sign in the bracket of Eq.~(\ref{loss_f}), it can be seen that these three equations are mutually restrained. We can directly get the adversarial properties:
Training $\theta _d$ to minimize $L_d$ (the domain loss) means $G_d$ cannot identify which domain the data comes from;
In the same way, minimizing $L_y$ by training $\theta _y$ is to predict the labels with high accuracy.
But it should be reminded that both $L_d$ and $L_y$ depend on the parameters of the feature extractor $\theta _f$,
and $\theta _f$ is determined by optimizing (minimizing) $ L(\theta _f,\theta _y,\theta _d)$ as an adversarial process. For a detailed explanation, please refer to  \cite{ganin2016domain,huembeli2018identifying,shen2021transfer}.
The network in this paper is implemented with TensorFlow-CPU 1.14 and Python 3.6.13 on Intel Xeon E5-1620 v4 CPU platform
with 16GB memory and 234GB storage space.

\subsection{Data sets of models}
\label{dataset}
Before feeding the configurations into DANN, they need to be partly labeled as source domain. 
To minimize the human intervention, we start to label the temperature ranges far away from the critical regime of the phase transition, where one can be sure that $T\ll T_c$ is an ordered, and $T\gg T_c$ is a disordered phase, as shown in Sec. \ref{Glauber}.
Therefore, we choose the configuration samples in $T\in [0.05, 0.1]$ below the phase transition point having label ``0" and $T\in [3.9, 4]$ above the phase transition point having label ``1" as the initial source domain, and the rest is taken to the target domain. 
Next, we are to narrow the target domain range, closing to the transition point, but keeping an "appropriate distance" from it. That means, that we set the new boundaries (temperature range) of the source domain range, and check with the network, that on that boundary at least 99\% of the predicted labels are the same.
Because the transition region is usually sharp for a system with enough large size, very few mislabeled samples can be negligible through automatic classification of the network.

After the training process of epoch = 1000, we use the target domain configurations to predict the classification of a configuration at each value of $T$ and average them for each $T$, separately. The output of DANN is a two-dimensional vector, which represents the two probabilities that the configurations (input data) belongs to the label ``0'' (ordered phase) and ``1'' (disordered phase). The critical point $T_c$ is defined as where the probabilities of the ordered and disorder phases are 50\%-50\%.

\subsection{The optimal source domain of DANN}\label{optimal}

To obtain more accurate predictions, we can reasonably expand the source domain with label to obtain more relevant information. As the initial source domain is $T\in[0.05, 0.1] \cup [3.9,4]$, the iterative method of interval expansion starts from $T\in[0.05, l^{(0)}] \cup [r^{(0)},4]$ (where $l^{(0)}=0.1$ and $r^{(0)}=3.9$) as follows \cite{shen2021transfer}:
\begin{equation}
l^{(i+1)}=\frac{l^{(i)}+T_{c,q}^{(i)}}{2},\quad 
r^{(i+1)}=\frac{r^{(i)}+T_{c,q}^{(i)}}{2},
\label{equ.lr}
\end{equation}
where $i$ represents the $i$-th expansion. $T_{c,q}^{(i)}$ is the estimate critical temperature by DANN on the $i$-th source domain interval. During each iteration, we need to check whether the output part of the range of source domain is at least 99\% in phase ``0'' or ``1'', as explained in the previous subsection. 
If so, the iteration process continues; otherwise, it is moved a step back to the initial value, for example, $l^{(i+1),1} \to (l^{(i+1),0})/2$. 
The correction is done if the confidence condition is not met and the process is stopped when it is no longer possible to expand the source domain. The effect of this part on $q=3$ can be seen in Fig.~\ref{Fig.q3.2}.

\section{Results}\label{sec:Results}

\subsection{Monte Carlo (MC) results }
\label{MC}

\begin{figure}[htbp]
\centering
\subfigure[]{
\label{Mag}
\includegraphics[height=0.3\textwidth]{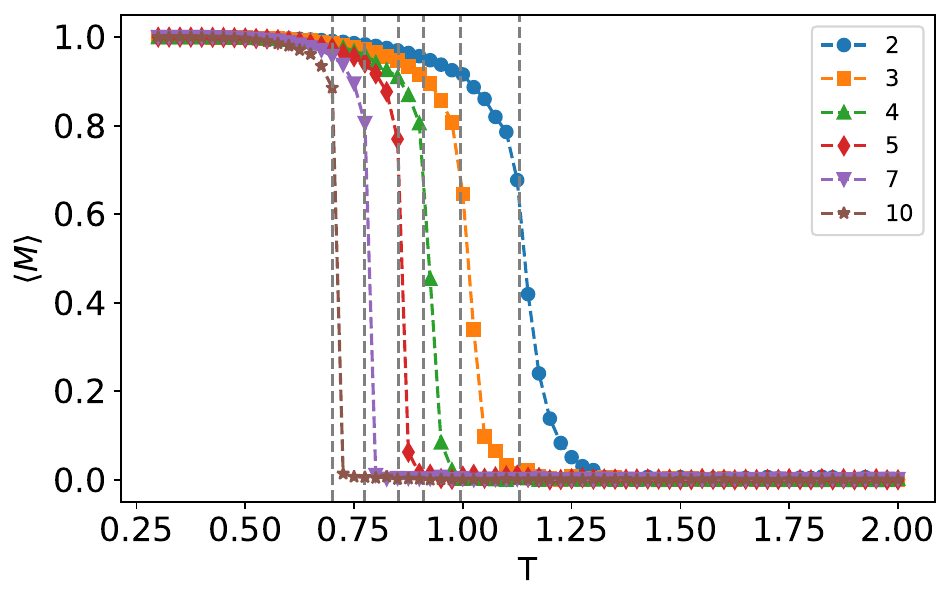}}
\subfigure[]{
\label{Ene}
\includegraphics[height=0.3\textwidth]{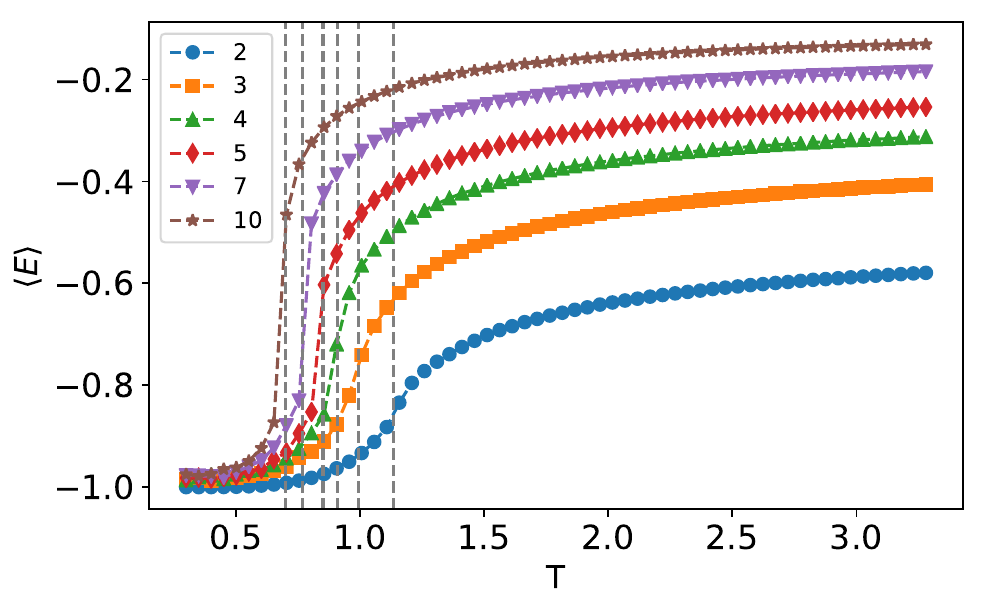}}
\caption{ The dependence of (a) magnetization $\langle M\rangle$ and (b) energy $\langle E \rangle$ on the temperature.  The dashed lines represent the theoretical critical points of each state. }
\label{fig:M}
\end{figure}

Even though we mainly use DANN,
we will begin the study of Potts model using the MC simulations (in order to verify the reliability of the above algorithm mechanism).
To investigate the nature of the phase transition, we performed MC simulations for $q$-state Potts model with $q = 2, 3, 4 , 5, 7$ and $10$. 

In previous studies, various variables were measured through the MC simulations approach to analyze criticality (phase transitions).
For instance, the magnetization (per site) $\left| \langle M\rangle \right|$, energy  $\langle E\rangle$,  specific heat $C_\nu$, magnetic susceptibility $\chi_s$    \cite{miyajima2021machine}.
Regardless of the up-down symmetry in the Potts model, an order parameter $\langle M\rangle$ can be defined \cite{henkel2008non}

\begin{equation}
\langle M\rangle=\frac{1}{N_s} \frac{1}{(q-1)L^d} \sum _{n=1}^{N_s} \sum _{i=1}^{L^d} [q\delta_{\sigma_{i,n},1 }-1] ,
\end{equation}
where $d=2$ and $L^d=L \times L$,
$N_s$ is the number of samples,
$\sigma_{i,n}$  represents the spin value at site $i$ of sample number $n$.

The energy of the system is given by
\begin{equation}
\langle E\rangle=\frac{1}{N_s} \frac{1}{L^d}  \sum _{n=1}^{N_s} \sum _{i=1}^{L^d} 
 \varepsilon \,,
\end{equation}
where  $\varepsilon$ is the energy defined in Eq.~(\ref{eq:energy}), and $\langle E \rangle$  is the mean energy per site averaged over $N_s=10000$ independent runs, with lattice size $L \times L$.

Fig.~\ref{fig:M} shows  $\langle M\rangle$ and $\langle E \rangle$ with  respect to  temperature for the $q$-state Potts model,
determined at  $L = 120$, respectively. 
In Fig.~\ref{Mag}, the  magnetization goes smoothly at the critical point for $q=2, 3, 4$ but has a clear  jump at $T_c$ for $q=5, 7, 10$.
Similar observation may be made for the energy $\langle E \rangle$ in Fig.~\ref{Ene}: a smooth transition for $q=2, 3, 4$, and a jump for $q>4$.

In short, it can be found that when $q \leq 4$, it is a second-order phase transition (continuous phase transition), and for $q >4 $, it is a first-order phase transition (discontinuous phase transition) \cite{wu1982potts,baxter1973potts}.
 There is an obvious phase transition at $T_c$  between the high temperature phase and the low temperature phase, (this  gives us confidence to proceed with ML),
which allows us to try to use ML (DANN) to distinguish the different phases according to the different configuration types at each temperature $T$ and therefore provides an alternative venue in addition to the Monte Carlo method. The advantage of such approach, that one has not to know a priori, how to calculate an order parameter, even does not need to know, what is the order parameter in the system: the network learns to identify the configurations with phases automatically.

\subsection{The DANN results of q=3}

\begin{figure}[htbp]
\centering
\subfigure[]{
\label{Fig.q3.1}
\includegraphics[height=0.3\textwidth]{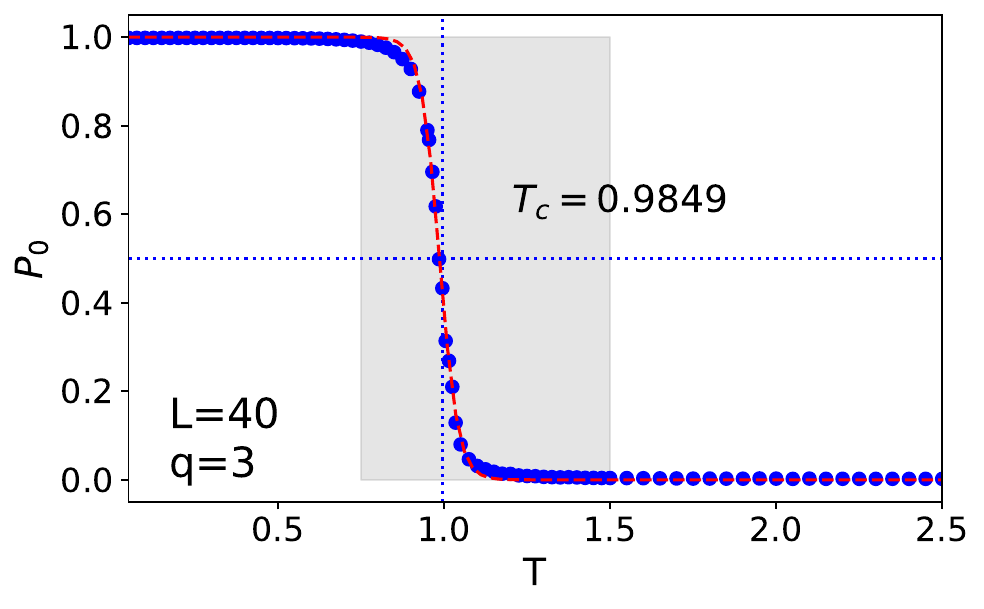}}
\subfigure[]{
\label{Fig.q3.2}
\includegraphics[height=0.3\textwidth]{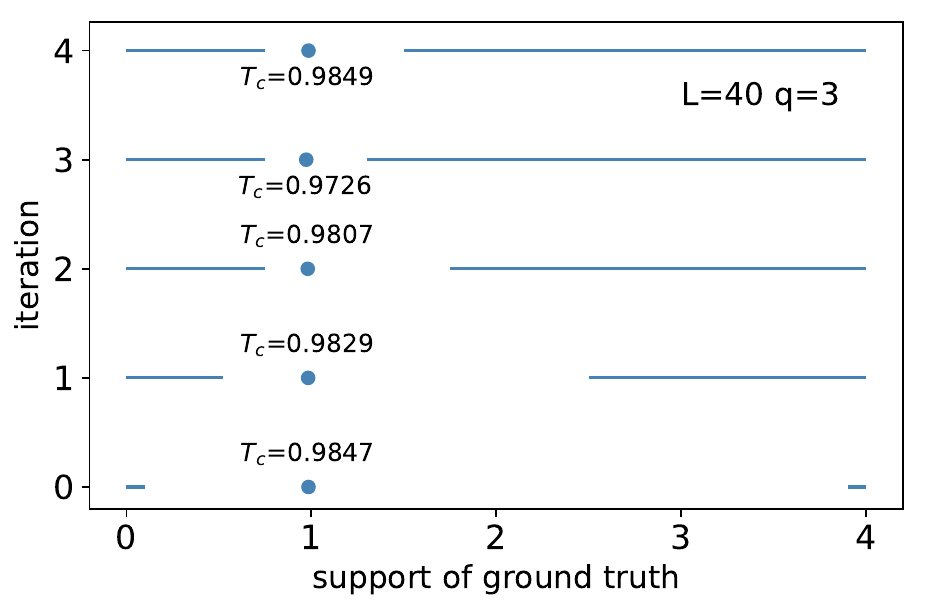}}
\caption{(a) Average probability of belonging to phase ``0'' for two-dimensional $3$-state Potts mode at $L = 40$, as a function of $T$. The shadowed region indicates the target domain at the optimal support, and the dashed line (red in color figure) is the sigmoid Eq.~\ref{eq:sigmoid}, fitting of the data. Its position parameter defines the critical temperature $T_c$. (b) Evolution of the optimal domain, and the corresponding critical temperatures. The x-axis is temperature $T$, and the y-axis represents the i-th iteration. Each blue line covers the range of the source domain, where the samples of chosen $T$ on the left interval are marked as ``0" and as ``1" on the right; The blue dots of $T_c$ is the estimated critical temperature by DANN on the i-th source domain interval.}
\label{fig:q3Tc}
\end{figure}

Firstly, we apply DANN on the Potts model with state $q=3$. Fig.~\ref{Fig.q3.1} shows the average probability $P_0$ belonging to phase ``0" at $L = 40$, obtained by training the DANN algorithm in the optimal source domain. By classifying samples at different temperatures $T$ into ``0" and ``1", the DANN returns a probability that a configuration belongs to ``0". The image of $P_0$ also can be fitted with a sigmoid function\footnote{we used this 1-sigmoid for the fits and in the figures.},
\begin{equation}
 T \to 1 - \frac1{1+e^{\frac{-(T-T_c)}{\sigma}}} \,,
 \label{eq:sigmoid}
\end{equation}
as indicated by the red dash line. For $P_0 =1/2$, we can get the corresponding $T$ as the critical temperature $T_c = 0.9849$. The final target domain at the optimal 
support is shown by the shadowed region, and Fig.~\ref{Fig.q3.2} gives the process of iteration method in detail. The optimal source domain can be reached within a few steps of iteration from the initial range $T\in[0.05, 0.1] \cup [3.9,4]$.

\begin{figure}[htbp]
\centering
\subfigure[]{
\label{Fig.q3.3}
\includegraphics[height=0.3\textwidth]{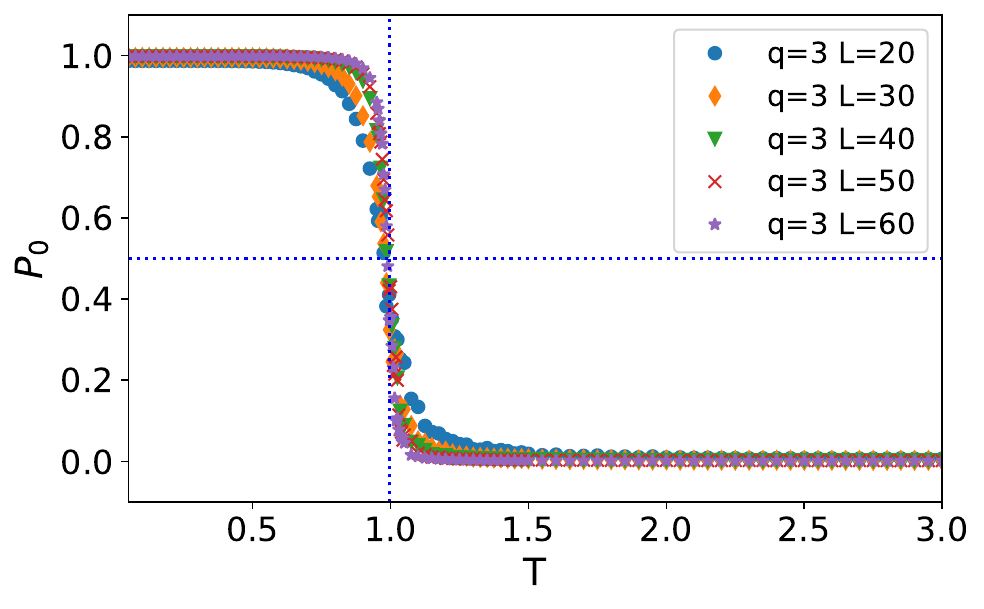}}
\subfigure[]{
\label{Fig.q3.4}
\includegraphics[height=0.3\textwidth]{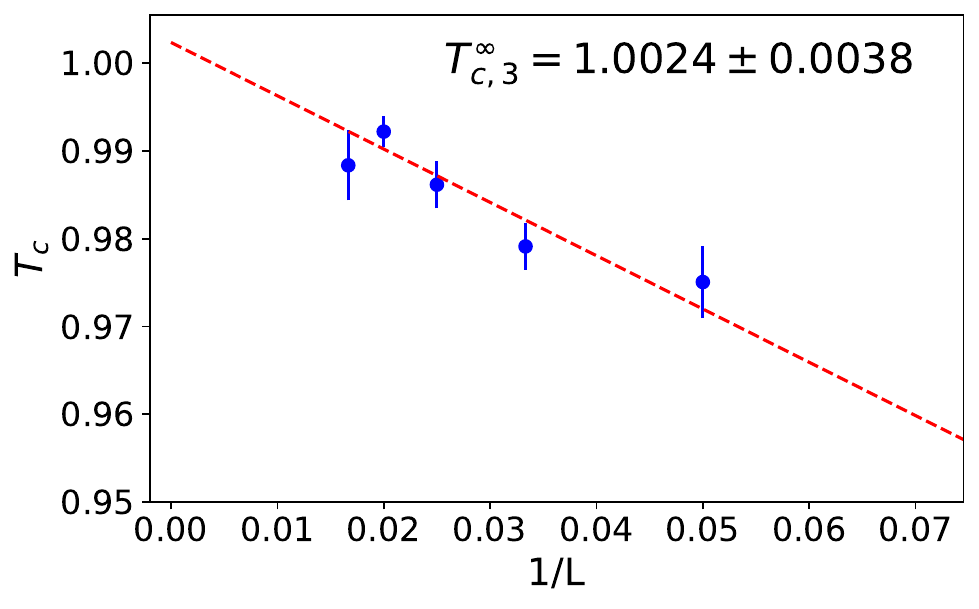}}
\caption{(a) Average probability of belonging to phase ``0" for two-dimensional $3$-state Potts model at $L = 20,30,40,50$ and $60$. (b) Extrapolation of the critical temperature $T_{c,q=3}$ to infinite lattice size is done with the assumption of linear dependence on $1/L$. The errorbars indicate the standard deviations  
calculated from $5$ independently trained neural networks.
}
\end{figure}

Repeating the procedure for
$L= 20, 30, 40, 50$ and $60$ as shown in Fig.~\ref{Fig.q3.3},  the critical point $T_{c,q=3}$  corresponding to infinite lattice size can be extracted by finite-size scaling (FSS) theory \cite{barber1983finite,fisher1972scaling,
privman1990finite}, extrapolating these results to zero on the $1/L$ scale with a linear fitting, as shown in Fig.~\ref{Fig.q3.4}.
At each size $L$ we run 5 independent trainings of the neural networks for the same dataset, obtaining an ensemble of critical temperatures. We use the average value for finite size scaling, and also estimate the errors $\xi_i$'s of the obtained temperature from the standard deviation. Hence, a weighted linear regression would be applied with weights, 
\begin{equation}
  w_i = \frac{\frac1{\xi_i^2}}{\sum_i \frac1{\xi_i^2}} \,.
\end{equation}
In the following we use the weighted linear regression, however, in Table~\ref{tab:NumericalResults} we also quote results obtained with uniform weights (unweighted linear regression).

After fitting, the critical value $T_{c,q=3}^{\infty}$ at $L \thicksim \infty$ is $1.0024  \pm 0.0038 $,  which is reasonably consistent with the MC simulations of Fan et al. \cite{fan2007determination} ( $T_c=0.9968(20)$ ), Ghaemi et al. \cite{2004Calculation} ( $T_c=0.9950(1)$ ), as well as the theoretical value $\frac{1}{\ln{(1+\sqrt{3})}}$ $\approx 0.9950$ \cite{wu1982potts}. The
\begin{equation}
  \chi^2/DOF  = \frac1{n-2} \sum_{i=1}^n \left( \frac{y_i - y_{pred,i}}{\xi_i} \right)^2
\end{equation}
value of the linear fit is 1.377.

\begin{figure}[htbp]
\subfigure[]{
\label{Fig.q3.5}
\includegraphics[height=0.33\textwidth]{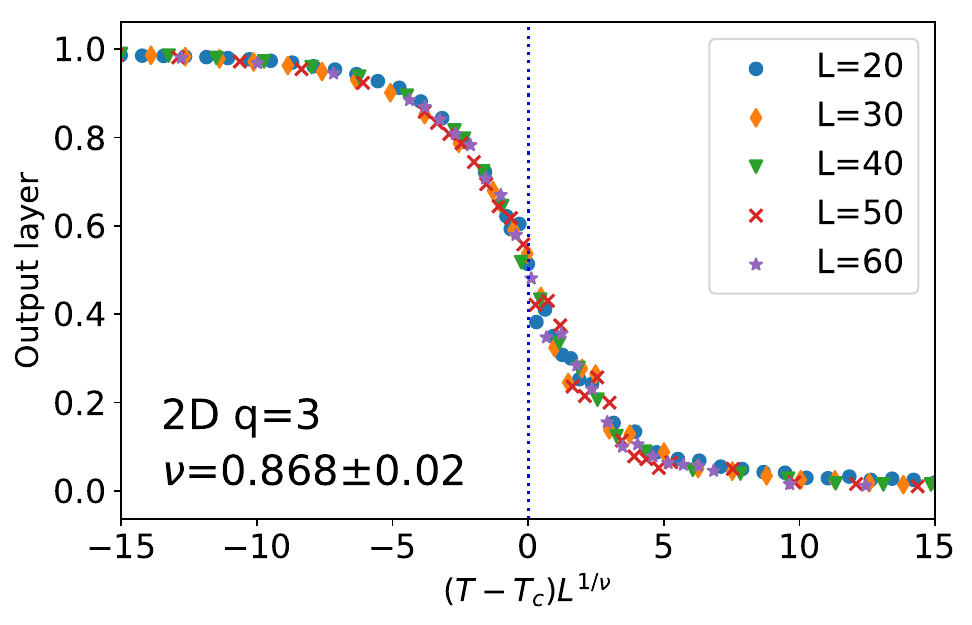}}
\subfigure[]{
\label{Fig.q3.6}
\includegraphics[height=0.33\textwidth]{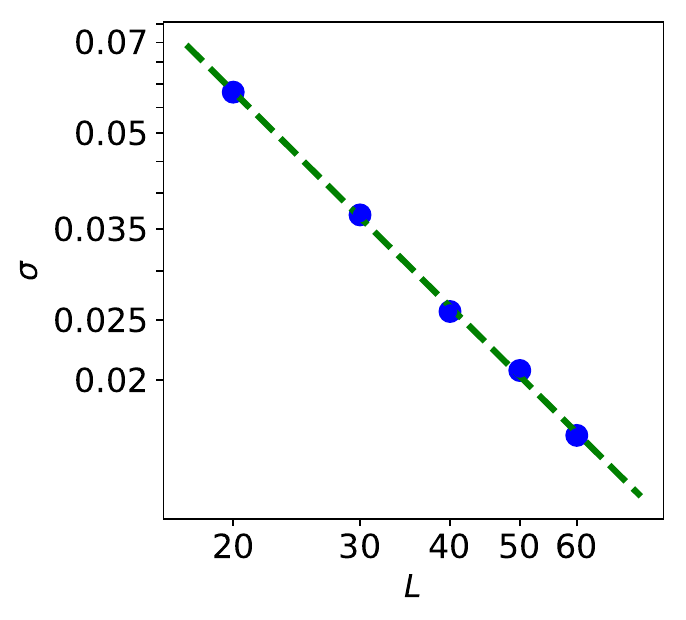}}
\caption{(a) Data collapse of $P_0$ as a function of $(T-T_c)L^{1/ \nu}$ for different system sizes. (b) Fit of the critical exponent $\nu$, using the scaling of the width $\sigma$, of the sigmoid (see Eq.~(\ref{eq:sigmoid}))  as a function of the system size $L = 20, 30, 40, 50, 60$. }
\label{fig:q3v1}
\end{figure}

Since all the results of DANN can be well fitted with a sigmoid function Eq.~(\ref{eq:sigmoid}),
we may also obtain the order parameter $\nu$ for continuous phase transition by a data collapse process.
This is cryptic:  Using the sigmoid parameters, we may plot $P_0$ as the function of $(T-T_c)/\sigma$. With this scaling the curve are collapsing into each other. Using the definition of the $\nu$ critical exponent, a scaling of the form $(T-T_c)L^{1/\nu}$ should hold, from which the critical exponent can read out as the slope of log-log fit in Fig.~\ref{Fig.q3.6}, and found to be
$\nu \simeq 0.868 \pm 0.02$, which is fairly  close to the theoretical value $5/6$ in the $3$-state Potts model \cite{wu1982potts,den1979relation}.  

\subsection{The DANN results of q=4}
\begin{figure}[htbp]
\centering
\subfigure[]{
\label{Fig.q4.1}
\includegraphics[height=0.3\textwidth]{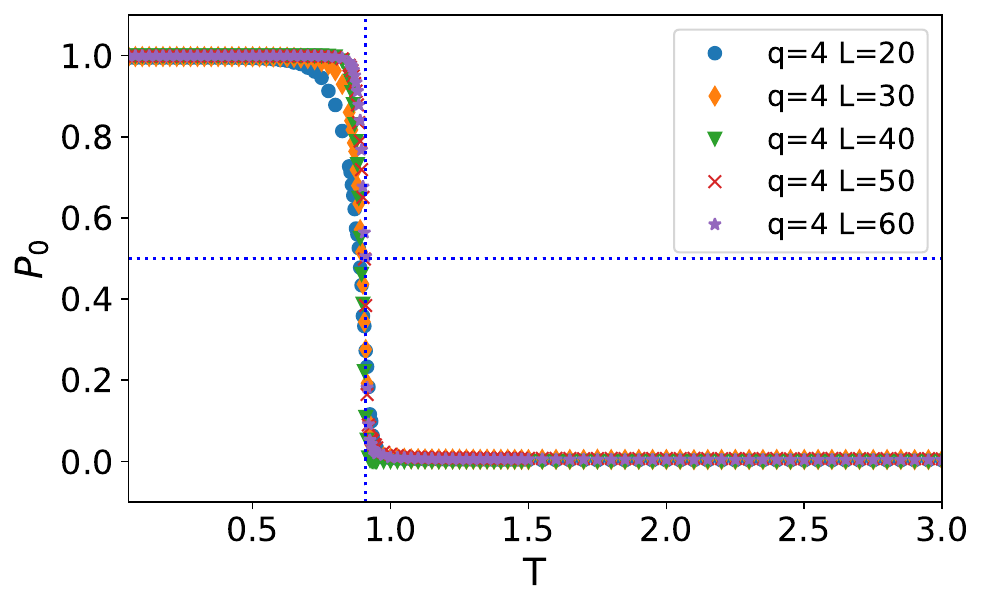}}
\subfigure[]{
\label{Fig.q4.2}
\includegraphics[height=0.3\textwidth]{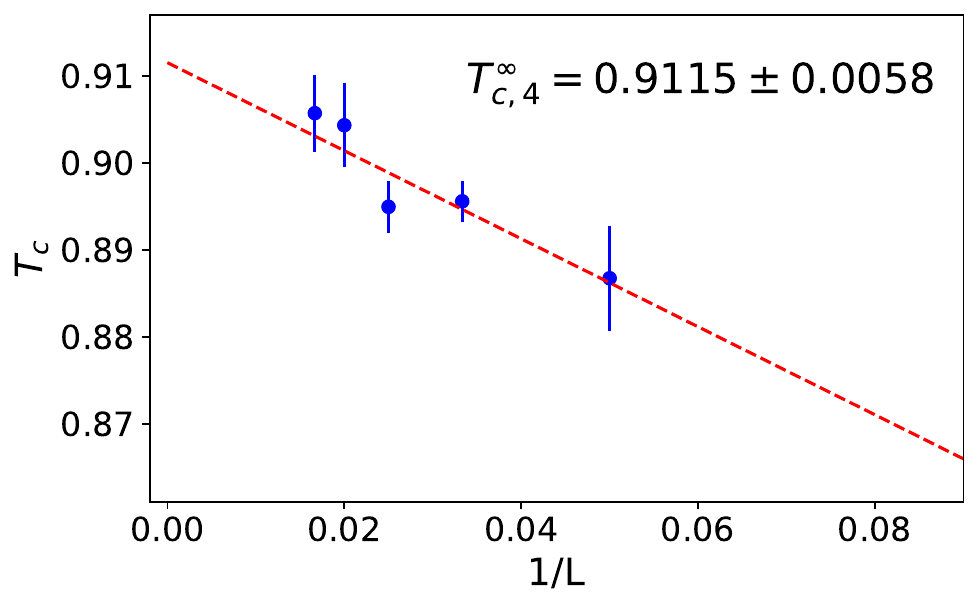}}
\caption{(a) Average probability of belonging to phase ``0" for two-dimensional $4$-state Potts model at $L = 20,30,40,50$ and $60$. (b) Extrapolation of the critical temperature $T_{c,q=4}$ to infinite lattice size by using FSS.}
\label{fig:q4crosspoint}
\end{figure}

Similarly to $q=3$, we apply DANN to the two dimensional $4$-state Potts model,  resulting in average probability $P_0$, as shown in Fig.~\ref{Fig.q4.1}, and
the extrapolation of the critical temperature to the infinite system in Fig.~\ref{Fig.q4.2}.
For the later, we found
$T_{c,q=4}^\infty=0.9115 \pm 0.0058 $, consistent with the theoretical result $\frac{1}{\ln{(1+\sqrt{4})}} \approx 0.9102$ and the MC simulation result in Ref.~\cite{hu1989monte} ($T_c=0.9104$). The $\chi^2/DOF$ value of the fit is 0.866.
Since two-dimensional $4$-state Potts model is known as a borderline case of first and second order phase transition, the leading power-law scaling behavior has to be modified by multiplicative logarithms \cite{salas1997logarithmic,cardy1986logarithmic}.

\begin{figure}[htbp]
\centering
\includegraphics[width=0.6\textwidth]{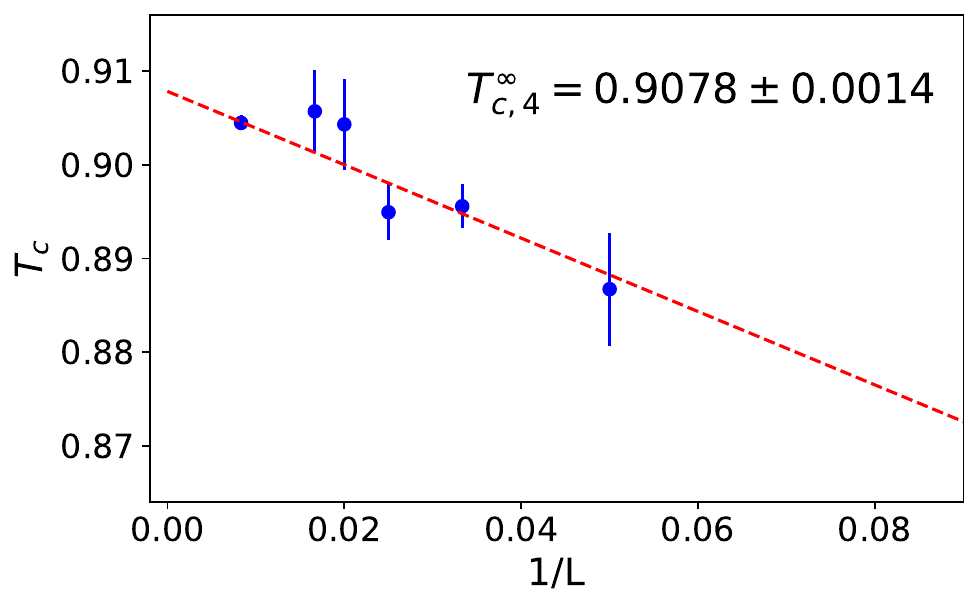}
\caption{Extrapolation of the critical temperature $T_{c,q=4}$ to infinite lattice size for $L=120$. }
\label{q4L120}
\end{figure}

For illustration, we performed the analysis for $L=120$ as well. The
critical temperature $T_c$ of $L = 120$ detected by DANN is $0.9045$, which is close to the
theoretical value $T_{c,q} = 0.9102$ and the fitted value. After adding the result of $L = 120$ for
extrapolation shown in Fig.~\ref{q4L120}, the new $T^{\infty}_{c,q=4} = 0.9078 \pm 0.0014$, is consistent with the
value $0.9115 \pm 0.0058$ in Fig.~\ref{Fig.q4.2}, and approaches to the theoretical one further. The $\chi^2/DOF$ value of the fit is calculated as $0.762$.

\subsection{The DANN results of q=5, 7, 10}	 
\begin{figure}[htbp]
\centering
\subfigure[]{
\label{Fig.q5.1}
\includegraphics[height=0.29\textwidth]{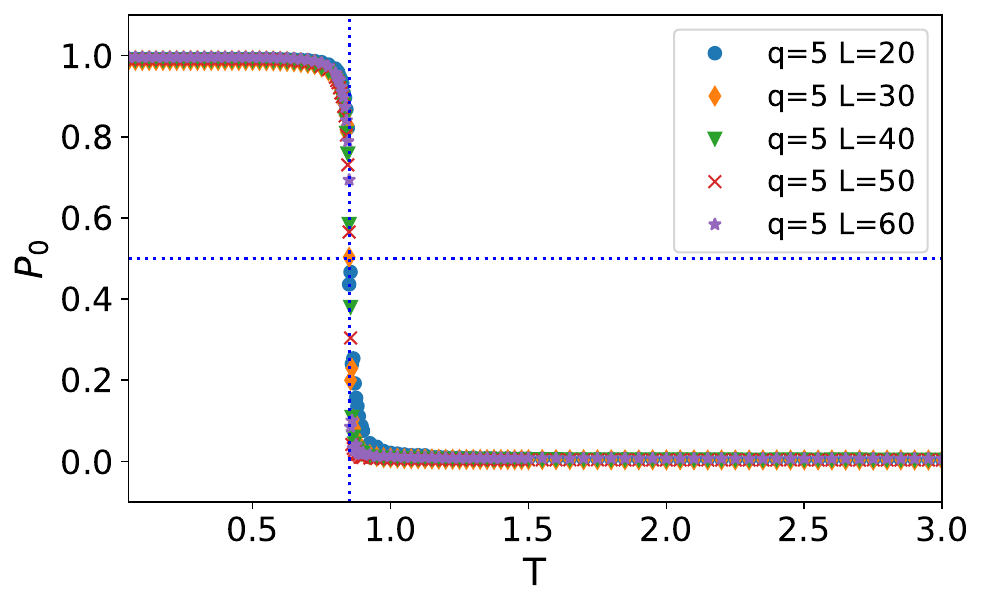}}
\subfigure[]{
\label{Fig.q5.2}
\includegraphics[height=0.29\textwidth]{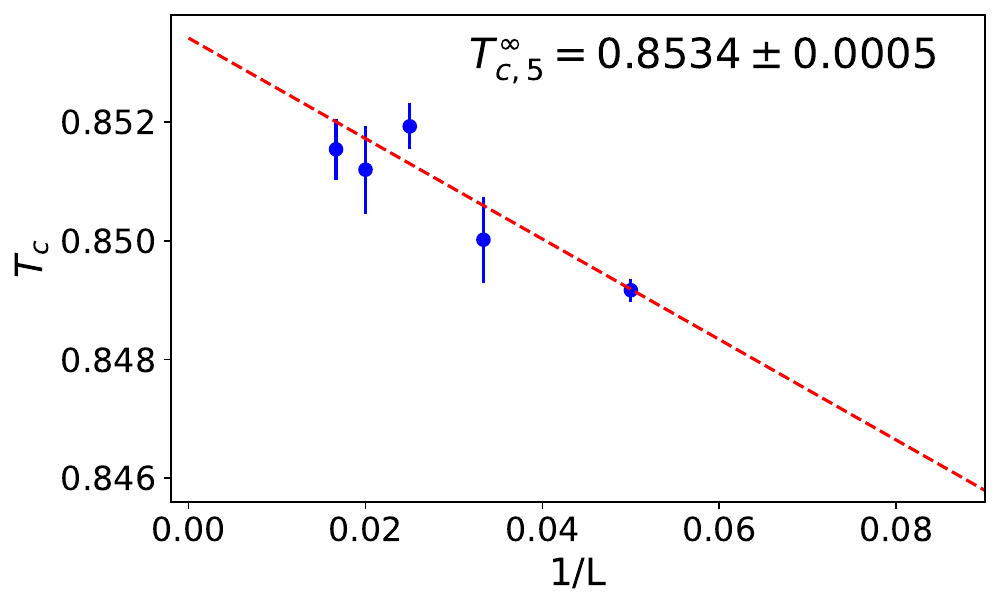}}
\subfigure[]{
\label{Fig.q7.1}
\includegraphics[height=0.29\textwidth]{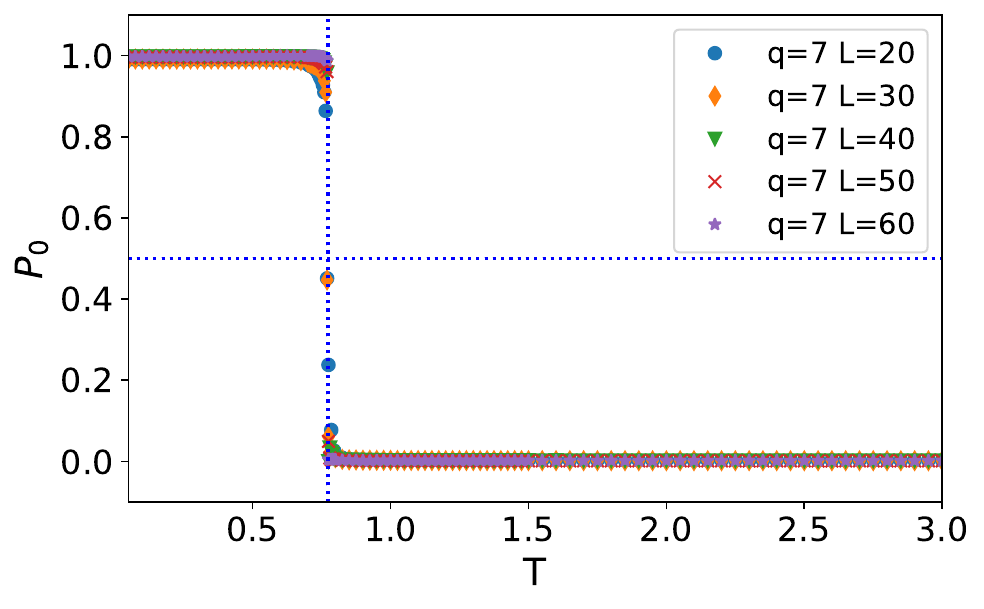}}
\subfigure[]{
\label{Fig.q7.2}
\includegraphics[height=0.29\textwidth]{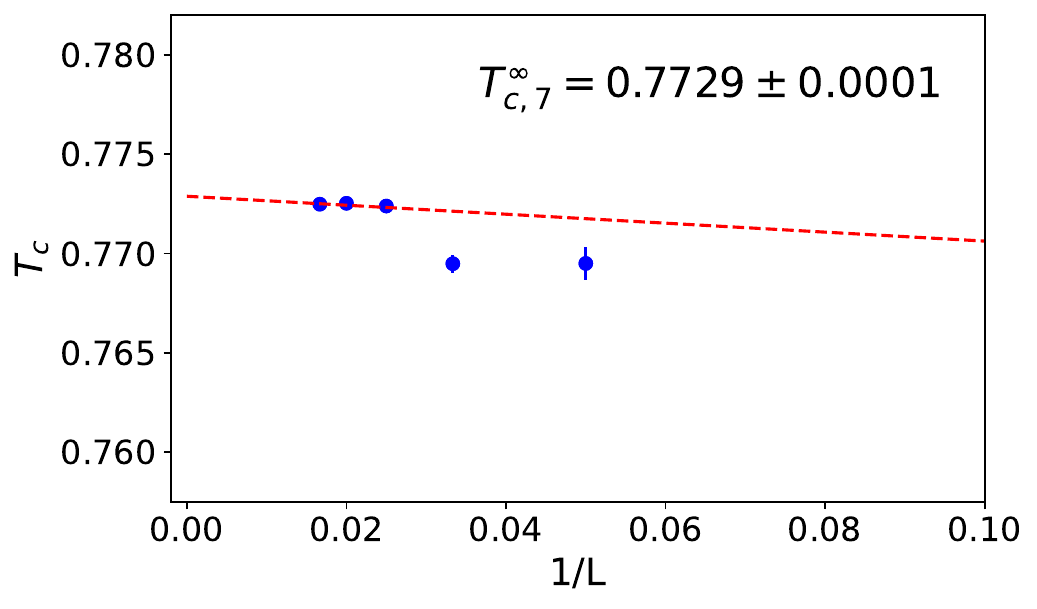}}
\subfigure[]{
\label{Fig.q10.1}
\includegraphics[height=0.29\textwidth]{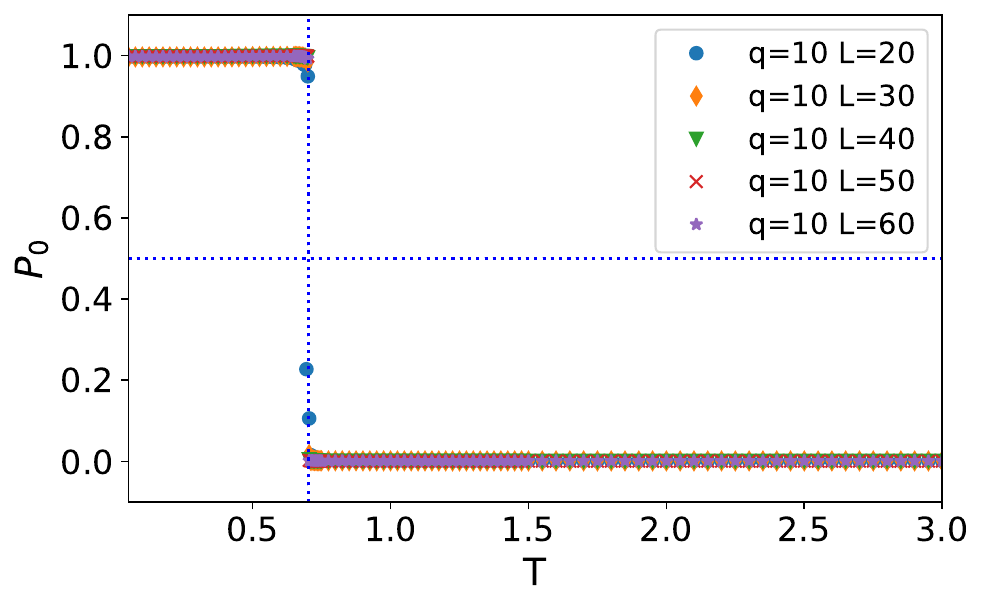}}
\subfigure[]{
\label{Fig.q10.2}
\includegraphics[height=0.29\textwidth]{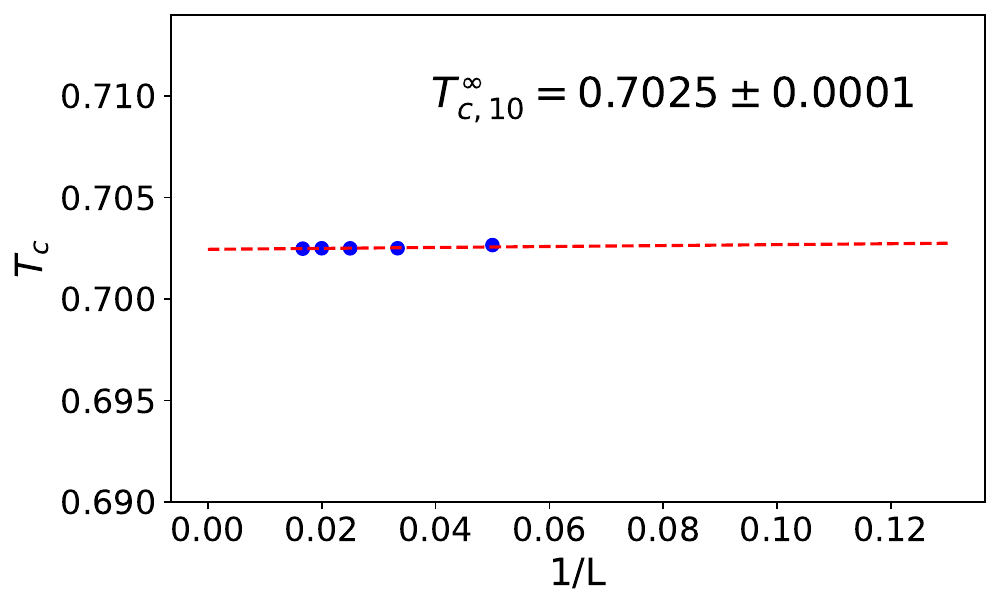}}

\caption{The average probability $P_0$ of two-dimensional Potts model for (a) $q= 5$, (c) $q=7$ and (e) $q=10$, with their extrapolation of the critical temperature to infinite lattice size (b) $T_{c,q=5}^\infty$, (d) $T_{c,q=7}^\infty$ and (f) $T_{c,q=10}^\infty$ by using FSS.}
\label{fig:q7q10}
\end{figure}

From the results of the two dimensional $3$- and $4$-state Potts model, it is successfully demonstrated the DANN can be applied to many-body systems.
For models with different $q$, the Potts model has different properties of phase transition and critical temperatures.
In order to make a more thorough analysis, we choose the $5$-,  $7$- and $10$-state Potts model additionally, and the DANN results are shown in Fig.~\ref{fig:q7q10}.

As the average probability $P_0$ of $q=5,7$ and $10$ shown in Fig.~\ref{Fig.q5.1}, \ref{Fig.q7.1} and \ref{Fig.q10.1},  we can clearly see that as $q$ increases, the transition is becoming increasingly abrupt near the critical temperature with a jump, in accordance with the theoretical predictions:
the phase transition is continuous for  $ q \leq 4$  
and discontinuous for $q>4$
\cite{wu1982potts,baxter1973potts}. 
The presence of a sudden numerical jump is an indication of the nature of phase transition, and a more thorough discussion is presented in Section~\ref{sec:det_trans}.

\begin{table}[htbp]
\caption{Critical properties of the $q$-state Potts model. $T_{c,q}$ is the theoretical value (see Eq.~(\ref{eq:Tc-theor})), $T_{c,q}^\infty$ is the critical temperature for infinite systems, and the transition types are first-order (1st) and second-order (2nd),
both predicted by DANN. For comparison we also present results obtained with unweighted linear regression.}
\label{tab:NumericalResults}
\centering
\begin{tabular}{lcccccccccc}
\hline\hline
q &$T_{c,q}$ \cite{wu1982potts} &$T_{c,q}^\infty$ (unweighted) &$T_{c,q}^\infty$ (weighted) &transition type&\\ \hline
3 &0.9950& 0.9983  $\pm$ 0.0072 & 1.0024  $\pm$ 0.0038 &2nd&\\
4 & 0.9102 & 0.9133 $\pm$ 0.0076 & 0.9115 $\pm$ 0.0058& 2nd & \\
5 &0.8515& 0.8530  $\pm$ 0.0014 & 0.8534  $\pm$ 0.0005&1st&\\ 
7 &0.7731& 0.7744  $\pm$ 0.0022 & 0.7729  $\pm$ 0.0001 &1st&\\
10 &0.7012& 0.7024  $\pm$ 0.0001 & 0.7025  $\pm$ 0.0001 &1st&\\
\hline\hline
\end{tabular}
\end{table}

Fig.~\ref{Fig.q5.2}, \ref{Fig.q7.2} and \ref{Fig.q10.2} show the extrapolation of the critical temperatures to the infinite system for $q=5,7$ and $10$. The obtained infinite size temperatures are summarized in Table.~\ref{tab:NumericalResults}, and they are all comparable with the theoretical results. We have found, that $T_{c, q}^\infty$ is more accurate for large $q$. In particular, the fitting curve of five different lattice sizes is almost a straight line to $y$ axis with $q = 10$ in Fig.~\ref{Fig.q10.2}.

The $\chi^2/DOF$'s for the fits are $1.536, 15.779, 1.483$ for $q=5,7$ and $10$, respectively. One may note the high value for $q=7$, due to the $N=20,30$ ``outliers" in Fig.~\ref{Fig.q7.2}. They are due to ripples (high uncertainty) of $P_0$ just above the critical temperature, where seemingly the statistic is not sufficient enough. While taking out the outliers, the $\chi^2/DOF$ reduces to $2.0$ for $q=7$ with $T_{c,q=7}^\infty = 0.7727\pm 0.0003$.

\subsection{The order of the phase transition}
\label{sec:det_trans}
 
To investigate the order of the phase transition (whether it is first, or second order one), we studied the distribution of the $P_0$'s (DANN predicted probabilities) of individual configurations around the transition temperature, similarly to the method described in~\cite{li2018applications,tan2020comprehensive}.

In case of a first order phase transitions one observers the coexistence of the two phases (droplets of the different phases), giving rise to a two-peaked distribution, one phase gradually overtaking as the temperature increases, while in the case of a second order phase transition this coexistence is missing due to the correlation length extending to the whole system.

\begin{figure}[htbp]
\centering
\subfigure[$q=3,T=0.965$]{
\label{Fig.configq3.1}
\includegraphics[width=28mm, height=23mm]{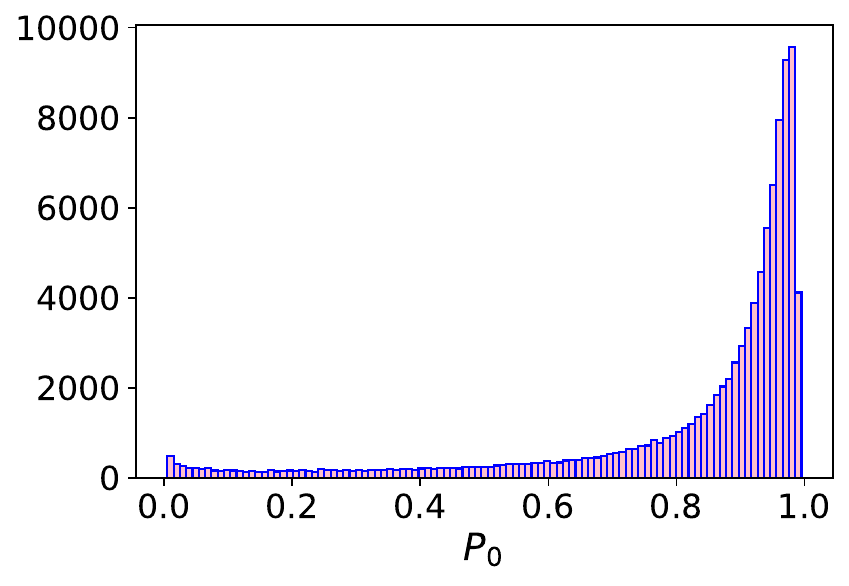}}
\subfigure[$q=3,T=0.975$]{
\label{Fig.configq3.2}
\includegraphics[width=28mm, height=23mm]{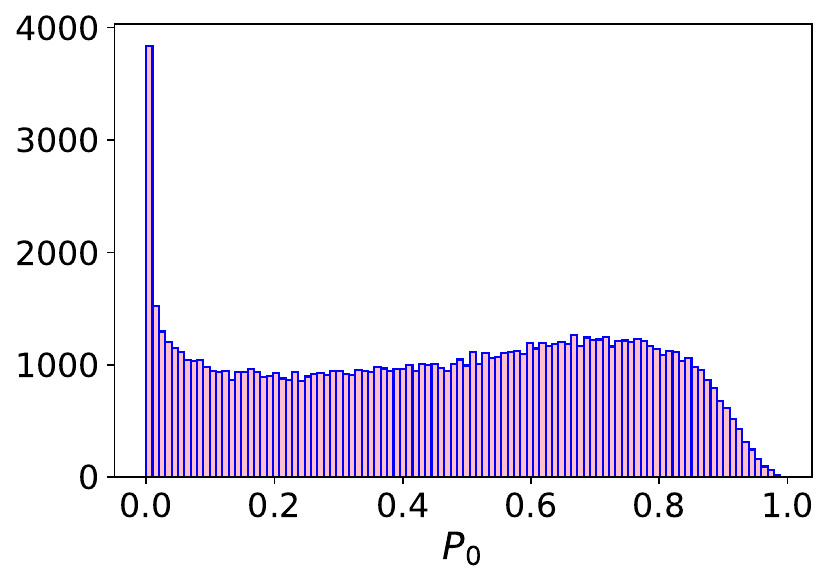}}
\subfigure[$q=4,T=0.875$]{
\label{Fig.configq3.3}
\includegraphics[width=28mm, height=23mm]{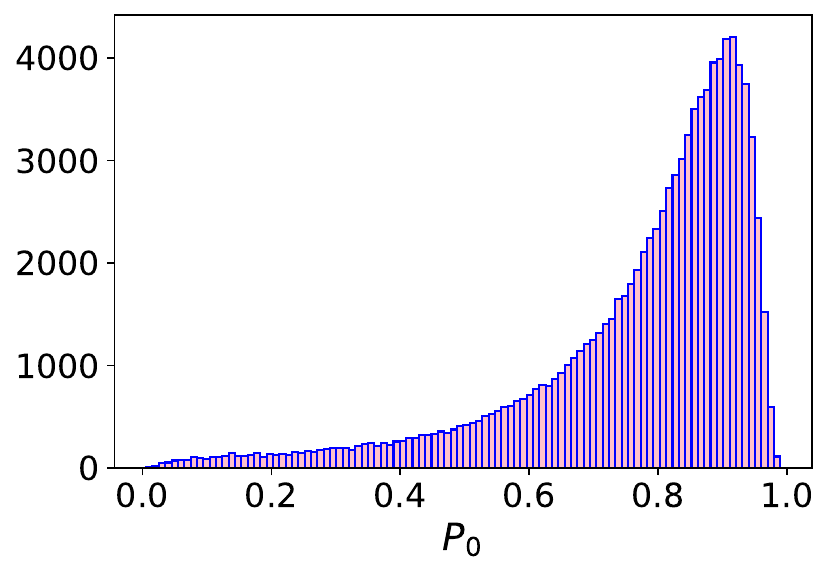}}
\subfigure[$q=4,T=0.886$]{
\label{Fig.configq3.4}
\includegraphics[width=28mm, height=23mm]{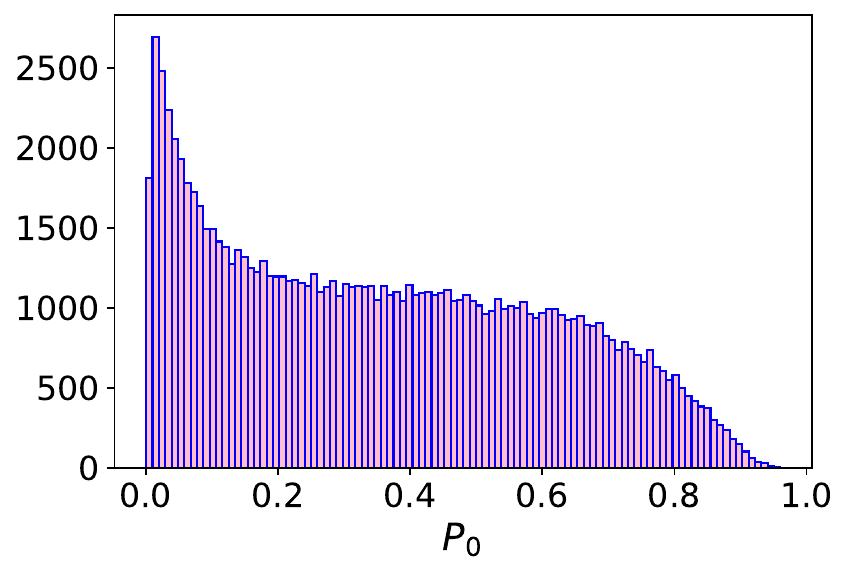}}
\subfigure[$q=5,T=0.848$]{
\label{Fig.configq3.5}
\includegraphics[width=28mm, height=23mm]{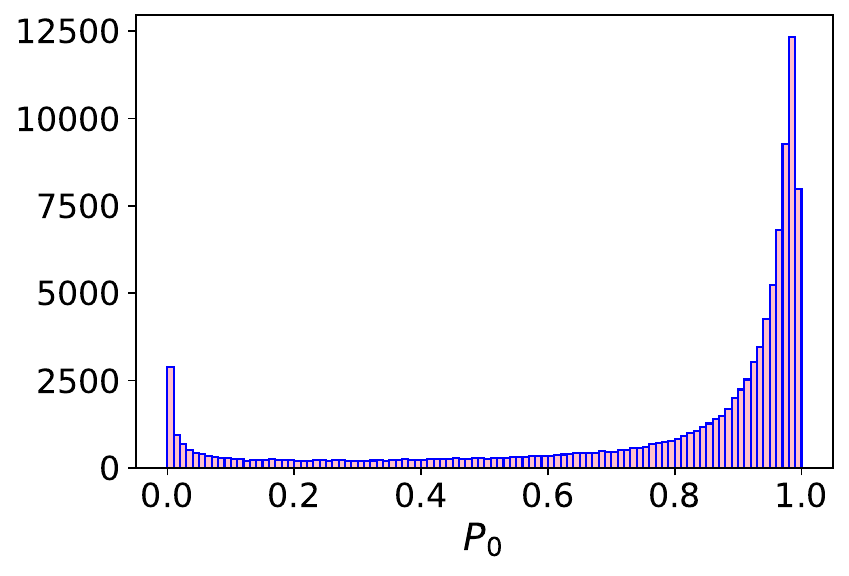}}
\subfigure[$q=5,T=0.85$]{
\label{Fig.configq3.6}
\includegraphics[width=28mm, height=23mm]{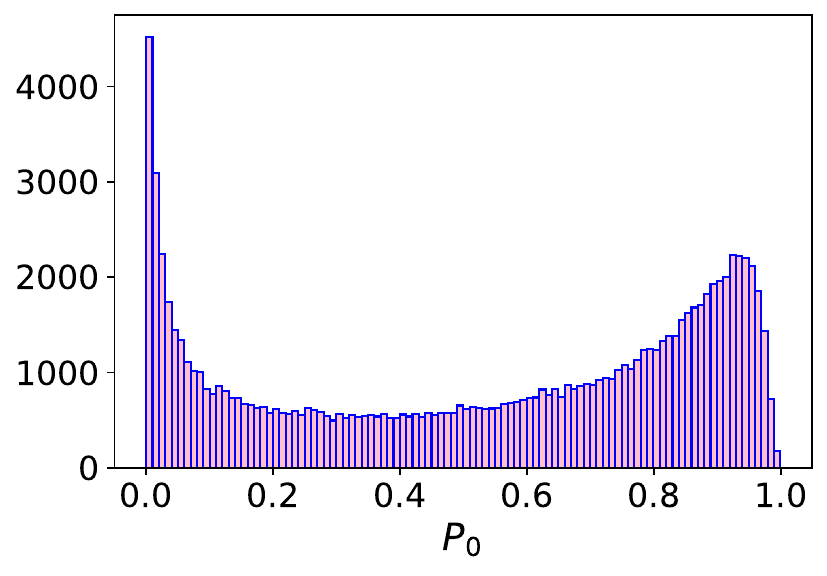}}
\subfigure[$q=7,T=0.77$]{
\label{Fig.configq3.7}
\includegraphics[width=28mm, height=23mm]{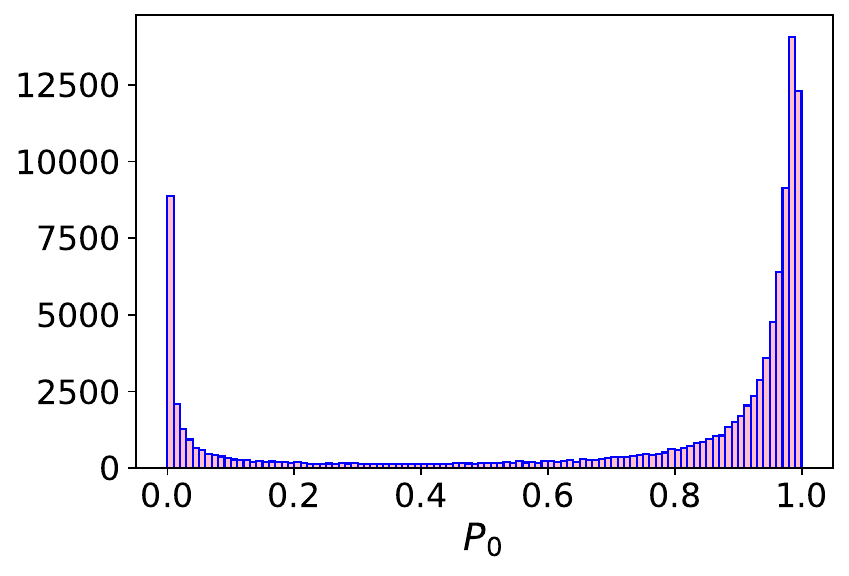}}
\subfigure[$q=7,T=0.771$]{
\label{Fig.configq3.8}
\includegraphics[width=28mm, height=23mm]{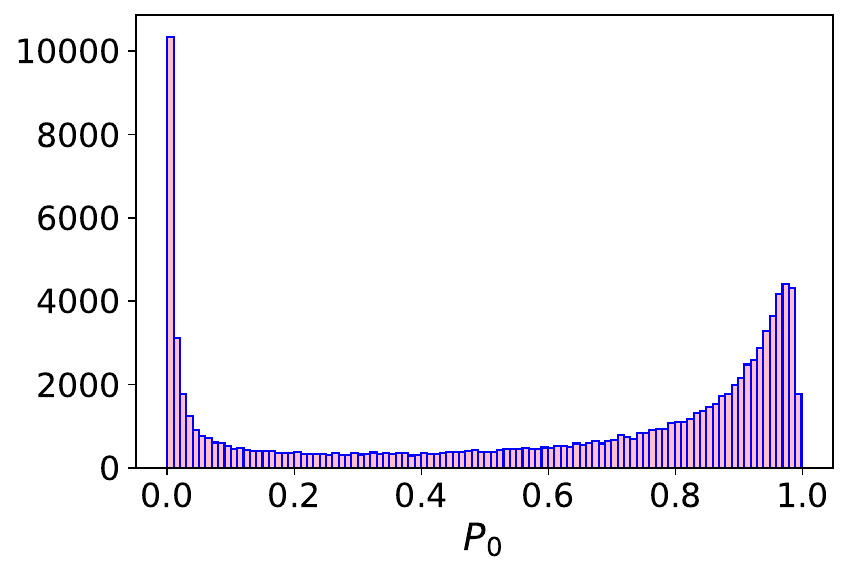}}
\subfigure[$q=10,T=0.7015$]{
\label{Fig.configq3.9}
\includegraphics[width=28mm, height=23mm]{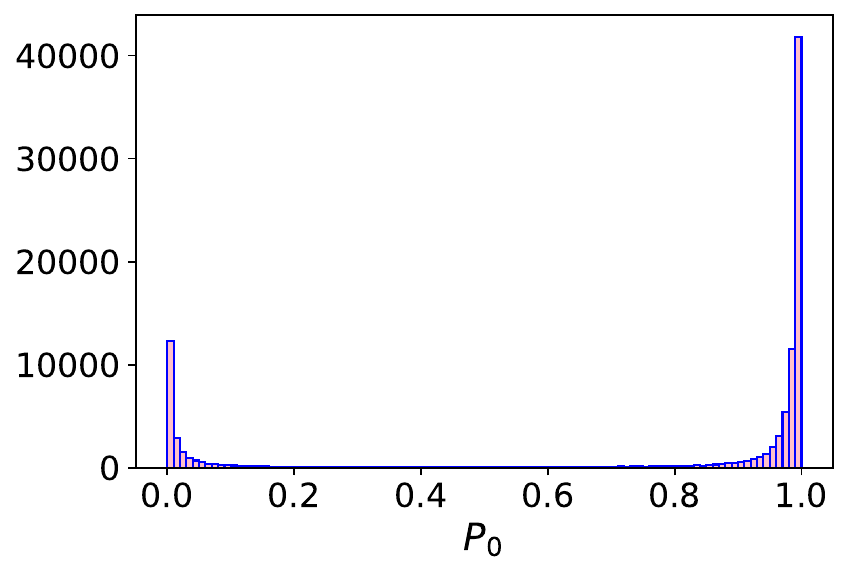}}
\subfigure[$q=10,T=0.7025$]{
\label{Fig.configq3.10}
\includegraphics[width=28mm, height=23mm]{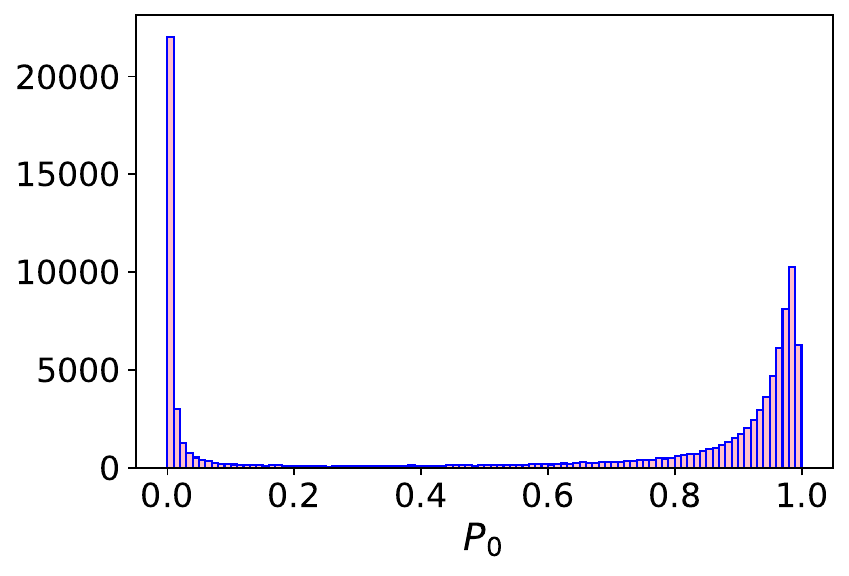}}

\caption{The histogram of DANN output $P_0 $ for each state with size $L=20$. $100000$ data points are generated for each of the plots, and each data point represents the output $P_0$ of a single configuration sample evaluated by DANN. }
\label{fig:DANNPdis}
\end{figure}

This is illustrated in Figs.~\ref{Fig.configq3.1}-\ref{Fig.configq3.10} where for each plot we have generated $10^5$ configurations both below and above $T_{c,q}$ for $q=3,4,5,7$ and $10$, for systems of size $L=20$, and plotted the distribution of DANN predicted probabilities of belonging to phase ``0", $P_0$. For $q\leq4$ the plots show always one peak, indicating a second order transition, while for $q> 4$ a two-peaked structure appears, a sign of a first order phase transition. As $q$ increases, the structure is becoming more pronounced.

To justify this result, we also performed the binning test, similarly to~\cite{li2018applications}, to check, whether the subsequent configurations are really independent. For $q=3$ and $q=4$ we averaged the predicted $P_0$ values over binsizes of 1, 2, 4, 8, 16, and 32, for $10^{5}$  configurations per size and $q$ values, and calculated the deviations of the averaged $P_0$'s for each bin size. In the presence of autocorrelation the deviation has a strong dependence on the bin size, while in our case we have observed no dependence on the bin size.

\subsection{Results of traditional NN and 2D CNN vs. DANN}\label{sec:Basics}

To test the efficiency and accuracy of the DANN, in this section we compare its results with a supervised learning and a 2D CNN approaches. Taking $q = 3$ at $L = 40$ as an example, for the supervised learning, we use the trained DANN algorithm for predictions, on the input set in the optimal source domain of the DANN. Here, all the configurations  below $T_c$ are labelled as ``0", and above $T_c$ as ``1".

\begin{figure}[htbp]
\centering
\label{Fig.cnn}
\includegraphics[height=0.35\textwidth]{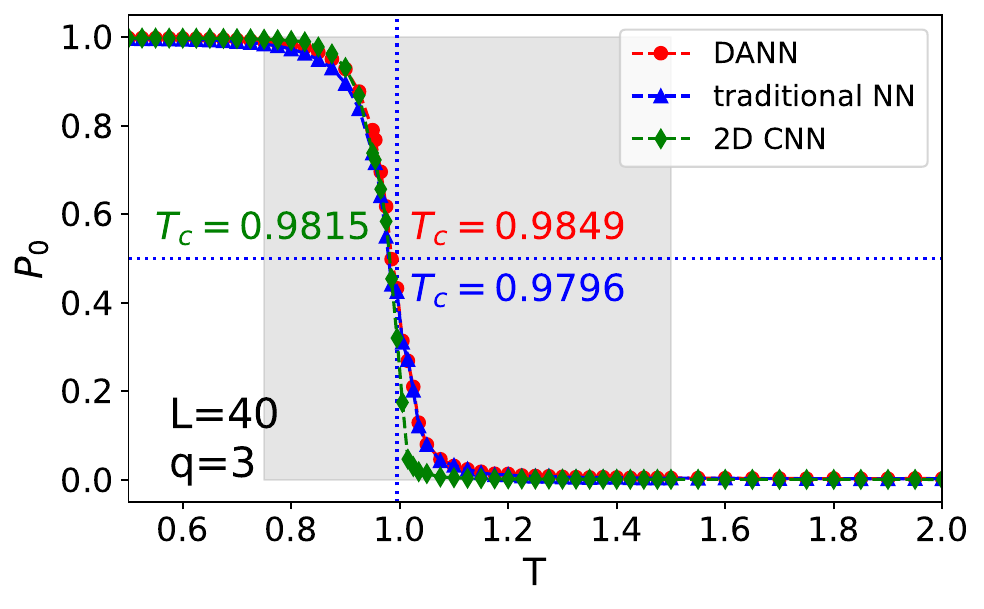}
\caption{DANN versus traditional NN and 2D CNN on $3$-state Potts model with $L=40$.}
\label{fig:cnn.nn}
\end{figure}

\begin{figure}[htbp]
\centering
\includegraphics[width=1\textwidth]{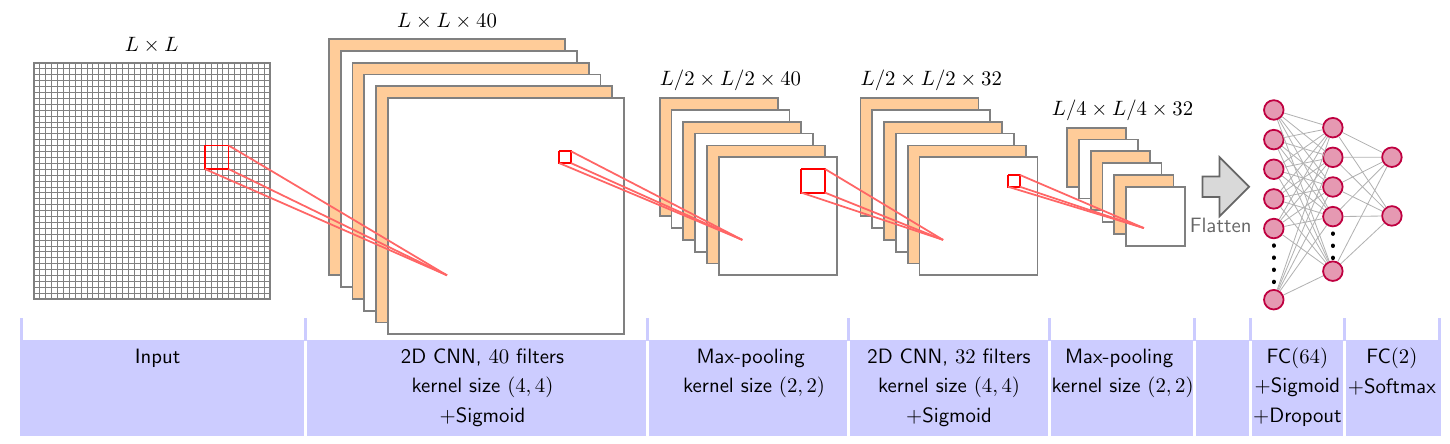}
\caption{Schematic structure of the neural network (a two-dimensional CNN), where the ``stride'' is set to the default value. The ``kernel size'' is the size of convolutional filter.}
\label{fig:CNN}
\end{figure}

Now, we compare the result from 3 models, the DANN, a 1D CNN (``traditional") model, being the original trained DANN model, but the domain classifier part subtracted~\cite{shen2021transfer} and a 2D CNN model (see Fig.~\ref{fig:CNN}). For the 2D CNN model we used two Max-pooling layers to increase the precision. The results are summarized in Fig.~\ref{fig:cnn.nn}: the critical temperature for the 1d CNN model is $0.9796$, smaller, than the full DANN result, $0.9849$. The 2D CNN yields $T_c = 0.9815$, close to the full DANN result. The 2D CNN training used 100 epochs, and the test was performed on 200 samples.
The results show, that the performance of the DANN is similar to the 2D CNN model, despite it was trained on unlabelled data.

Next, we compare the computation time of DANN and  supervised networks. Table.~\ref{tab:timecost} gives the cost time of the whole training and test process of the DANN and the supervised 2D CNN for $q = 3$ at $L = 20, 30, 40, 50$ and $60$. It can be seen that the required time for the algorithms increases with the lattice size. For the DANN, when $i$, the number of iterations for searching the optimal source domain is smaller than $8$, the time cost of the DANN is less than the 2D CNN. Usually, $i<6$, it takes only few iteration steps to find out the optimal source domain, like $q=3$ shown in Table.~\ref{tab:timecost}, so the DANN method is more efficient than traditional supervised network (2D CNN).

\begin{table}[htbp]
\caption{ The comparison of computational time (in terms of seconds) between the 2D CNN and DANN in $3$-state Potts model. $i$ indicates the number of iterations in the evolution of the optimal domain support for DANN and is $5$ for $q=3$. 
The running time for DANN should be approximated as the average time $\times$ the total iteration steps $i$.
}
\label{tab:timecost}
\centering
\resizebox{\linewidth}{!}{
\begin{tabular}{lcccccc}
\hline\hline
Lattice size &L=20&L=30&L=40&L=50&L=60&\\ \hline
Time cost (2D CNN)&2091.5s  & 3568.1s  &5573.2s  & 10473.8s  & 20536.3s  &\\
Time cost (DANN) &30.3 s $\times i$ & 51.4 s $\times i$ & 74.2 s $\times i$ & 102.8  s $ \times i$  &129.4 s $\times  i$ &\\
\hline\hline
\end{tabular}}
\end{table}

\section{Conclusion} 

In this paper, we investigate the phase transitions of two-dimensional $q$-state Potts model using Monte Carlo simulation and machine learning. The Monte Carlo simulation verifies the reliability of the Glauber algorithm~\cite{henkel2008non,glauber1963time,mariz1990comparative} and reproduces the properties of the first-order and second-order phase transitions. 
The particular focus here is machine learning, where 
we use a semi-supervised machine learning method, DANN, to predict the critical temperature of the two-dimensional $q$-state Potts model. It is a powerful method based on adversarial learning, where only part of the input data needs to be labeled and the remaining labels can be predicted. This property of machine learning allows us to estimate the critical temperature of phase transition only by a few labeled samples of configuration. For different $q$, the Potts model has different phase transition behaviors and critical temperatures, so we applied DANN on the Potts model with $q =3$ and $4, 5,7,10$ at lattice sizes $L = 20, 30, 40, 50$ and $60$. An iterative method is introduced to find the optimal source domain, which makes the label set information learned by training more complete and accurate. 

The output of the DANN is $P_0$, the average probability of belonging to phase ``0". For all $q$'s studied in this paper, $P_0$ as a function of the temperature, $T$ may be fitted with a sigmoid function. Temperature $T$, corresponding to $P_0 = 50\%$, defines the estimated critical temperature $T_c$.  Its value for the infinite lattice size can be obtained by extrapolating $T_c$ on the $1/L$ scale to $1/L=0$. The resulting values were found to be close to the theoretical ones, e.g.  $T_{c,q=3}^\infty = 1.0024  \pm 0.0038$ for $q=3$, is close to the theoretical value  $\frac{1}{\ln(1+\sqrt{3})}$. At the same time, we can also calculate the order parameter $\nu \simeq 0.868 \pm 0.02$ by a data collapse process, comparable to the theoretical value of $5/6$.

For $q=4$, similarly to $q=3$, we found a second order phase transition with critical temperature close to the theoretical one. 
For $q = 5, 7$ and $10$, $P_0$ has a jump as the function of the temperature in the region of phase transition, which is consistent with the behavior of the first-order phase transition in theory. The obtained critical temperatures are also consistent with the theoretical values and  we found them to be more accurate with increasing $q$.

Furthermore, we compared the results of DANN with traditional supervised learning, and showed that they are consistent.
Compared to the traditional supervised learning, the advantage of DANN for phase transition is its extraction power. The training set of traditional supervised learning needs to be fully labeled, that is, all the information of the training set is known. However, the training set for the DANN  needs only a small part to be labelled, and the rest can be predicted from this, which allows us to explore phase transition models that cannot be solved theoretically.  
We also found, that parallel to the accurate prediction of the critical temperature, using DANN it is also possible to distinguish first and second order transitions.

In the paper, we were concentrating on small size systems and showed that already with quite small systems DANN has good performance to capture the features of different
phases from unlabelled data. 
Accuracy may be further improved using larger lattice sizes, however, it requires more computation resources.

\section*{Acknowledgements}
We gratefully acknowledge the fruitful discussions with Shengfeng Deng, Dian Xu and Kui Tuo. 
This work was supported in part by National Natural Science Foundation of China (Grant No. 61873104, 11505071), the Programme of Introducing Talents of Discipline to Universities under Grant no. B08033, the Fundamental Research Funds for the Central Universities, and the European Union project 
RRF-2.3.1-21-2022-00004 within the framework of MILAB.


\bibliography{DANN}

\end{document}